%$Revision: 1.50 $
%$Date: 2007/09/25 14:38:21 $
\input harvmac

\input amssym
\input epsf
\input tables

\def\unit{\relax{\rm 1\kern-.26em I}}
\def\nada{\relax{\rm 0\kern-.30em l}}
\def\tilde{\widetilde}

%\draftmode

\def\figin{\epsfcheck\figin}\def\figins{\epsfcheck\figins}
\def\epsfcheck{\ifx\epsfbox\UnDeFiNeD
\message{(NO epsf.tex, FIGURES WILL BE IGNORED)}
\gdef\figin##1{\vskip2in}\gdef\figins##1{\hskip.5in}% blank space instead
\else\message{(FIGURES WILL BE INCLUDED)}%
\gdef\figin##1{##1}\gdef\figins##1{##1}\fi}
\def\DefWarn#1{}
\def\figinsert{\goodbreak\midinsert}
\def\ifig#1#2#3{\DefWarn#1\xdef#1{fig.~\the\figno}
\writedef{#1\leftbracket fig.\noexpand~\the\figno}%
\figinsert\figin{\centerline{#3}}\medskip\centerline{\vbox{\baselineskip12pt
\advance\hsize by -1truein\noindent\footnotefont{\bf
Fig.~\the\figno:\ } \it#2}}
\bigskip\endinsert\global\advance\figno by1}

\def\det{{\rm det}}

%% MACROS
\noblackbox
\def\IL{\relax{\rm I\kern-.18em L}}
\def\IH{\relax{\rm I\kern-.18em H}}
\def\IR{\relax{\rm I\kern-.18em R}}
\def\IC{\relax\hbox{$\inbar\kern-.3em{\rm C}$}}
\def\IZ{\relax\ifmmode\mathchoice
{\hbox{\cmss Z\kern-.4em Z}}{\hbox{\cmss Z\kern-.4em Z}}
{\lower.9pt\hbox{\cmsss Z\kern-.4em Z}} {\lower1.2pt\hbox{\cmsss
Z\kern-.4em Z}}\else{\cmss Z\kern-.4em Z}\fi}
\def\CM {{\cal M}}

\def\CO {{\cal O}}

%% MORE MACROS
\def\CM {{\cal M}}

\def\CO {{\cal O}}

\def\det{{\rm det}}

\font\manual=manfnt \def\dbend{\lower3.5pt\hbox{\manual\char127}}

\def\IZ{\relax\ifmmode\mathchoice
{\hbox{\cmss Z\kern-.4em Z}}{\hbox{\cmss Z\kern-.4em Z}}
{\lower.9pt\hbox{\cmsss Z\kern-.4em Z}} {\lower1.2pt\hbox{\cmsss
Z\kern-.4em Z}}\else{\cmss Z\kern-.4em Z}\fi}
\def\half {{1\over 2}}

\def\lfm#1{\medskip\noindent\item{#1}}

\def\bar{\overline}

\def\rt2{\sqrt{2}}
\def\irt2{{1\over\sqrt{2}}}

%  \slashchar puts a slash through a character to represent contraction
%  with Dirac matrices. Use \not instead for negation of relations, and use
%  \hbar for hear.
\def\slashchar#1{\setbox0=\hbox{$#1$}           % set a box for #1
   \dimen0=\wd0                                 % and get its size
   \setbox1=\hbox{/} \dimen1=\wd1               % get size of /
   \ifdim\dimen0>\dimen1                        % #1 is bigger
      \rlap{\hbox to \dimen0{\hfil/\hfil}}      % so center / in box
      #1                                        % and print #1
   \else                                        % / is bigger
      \rlap{\hbox to \dimen1{\hfil$#1$\hfil}}   % so center #1
      /                                         % and print /
   \fi}

\def\foursqr#1#2{{\vcenter{\vbox{
    \hrule height.#2pt
    \hbox{\vrule width.#2pt height#1pt \kern#1pt
    \vrule width.#2pt}
    \hrule height.#2pt
    \hrule height.#2pt
    \hbox{\vrule width.#2pt height#1pt \kern#1pt
    \vrule width.#2pt}
    \hrule height.#2pt
        \hrule height.#2pt
    \hbox{\vrule width.#2pt height#1pt \kern#1pt
    \vrule width.#2pt}
    \hrule height.#2pt
        \hrule height.#2pt
    \hbox{\vrule width.#2pt height#1pt \kern#1pt
    \vrule width.#2pt}
    \hrule height.#2pt}}}}
\def\psqr#1#2{{\vcenter{\vbox{\hrule height.#2pt
    \hbox{\vrule width.#2pt height#1pt \kern#1pt
    \vrule width.#2pt}
    \hrule height.#2pt \hrule height.#2pt
    \hbox{\vrule width.#2pt height#1pt \kern#1pt
    \vrule width.#2pt}
    \hrule height.#2pt}}}}
\def\sqr#1#2{{\vcenter{\vbox{\hrule height.#2pt
    \hbox{\vrule width.#2pt height#1pt \kern#1pt
    \vrule width.#2pt}
    \hrule height.#2pt}}}}

\lref\martinreview{
  S.~P.~Martin,
  ``A supersymmetry primer,''
  arXiv:hep-ph/9709356.
  %%CITATION = HEP-PH/9709356;%%
}

%\DineYW
\lref\DineYW{
  M.~Dine and A.~E.~Nelson,
  ``Dynamical supersymmetry breaking at low-energies,''
  Phys.\ Rev.\  D {\bf 48}, 1277 (1993)
  [arXiv:hep-ph/9303230].
  %%CITATION = PHRVA,D48,1277;%%
}

%\DineVC
\lref\DineVC{
  M.~Dine, A.~E.~Nelson and Y.~Shirman,
  ``Low-Energy Dynamical Supersymmetry Breaking Simplified,''
  Phys.\ Rev.\  D {\bf 51}, 1362 (1995)
  [arXiv:hep-ph/9408384].
  %%CITATION = PHRVA,D51,1362;%%
}

%\DineAG
\lref\DineAG{
  M.~Dine, A.~E.~Nelson, Y.~Nir and Y.~Shirman,
  ``New tools for low-energy dynamical supersymmetry breaking,''
  Phys.\ Rev.\  D {\bf 53}, 2658 (1996)
  [arXiv:hep-ph/9507378].
  %%CITATION = PHRVA,D53,2658;%%
}

\lref\GiudiceBP{
  G.~F.~Giudice and R.~Rattazzi,
  ``Theories with gauge-mediated supersymmetry breaking,''
  Phys.\ Rept.\  {\bf 322}, 419 (1999)
  [arXiv:hep-ph/9801271].
  %%CITATION = PRPLC,322,419;%%
}

%\WittenNF
\lref\WittenNF{
  E.~Witten,
  ``Dynamical Breaking Of Supersymmetry,''
  Nucl.\ Phys.\  B {\bf 188}, 513 (1981).
  %%CITATION = NUPHA,B188,513;%%
}

\lref\CohenQC{
  A.~G.~Cohen, T.~S.~Roy and M.~Schmaltz,
  ``Hidden sector renormalization of MSSM scalar masses,''
  JHEP {\bf 0702}, 027 (2007)
  [arXiv:hep-ph/0612100].
  %%CITATION = JHEPA,0702,027;%%
}

%\DimopoulosVZ
\lref\DimopoulosVZ{
  S.~Dimopoulos, M.~Dine, S.~Raby and S.~D.~Thomas,
  ``Experimental Signatures of Low Energy Gauge Mediated Supersymmetry
  Breaking,''
  Phys.\ Rev.\ Lett.\  {\bf 76}, 3494 (1996)
  [arXiv:hep-ph/9601367].
  %%CITATION = PRLTA,76,3494;%%
}

%\AmbrosanioZR
\lref\AmbrosanioZR{
  S.~Ambrosanio, G.~L.~Kane, G.~D.~Kribs, S.~P.~Martin and S.~Mrenna,
  ``Supersymmetric analysis and predictions based on the CDF ee$\gamma\gamma$ +
  missing $E_T$ event,''
  Phys.\ Rev.\ Lett.\  {\bf 76}, 3498 (1996)
  [arXiv:hep-ph/9602239].
  %%CITATION = PRLTA,76,3498;%%
}

%\DimopoulosVA
\lref\DimopoulosVA{
  S.~Dimopoulos, S.~D.~Thomas and J.~D.~Wells,
  ``Implications of low energy supersymmetry breaking at the Tevatron,''
  Phys.\ Rev.\  D {\bf 54}, 3283 (1996)
  [arXiv:hep-ph/9604452].
  %%CITATION = PHRVA,D54,3283;%%
}

%\AmbrosanioJN
\lref\AmbrosanioJN{
  S.~Ambrosanio, G.~L.~Kane, G.~D.~Kribs, S.~P.~Martin and S.~Mrenna,
  ``Search for supersymmetry with a light gravitino at the Fermilab  Tevatron
  and CERN LEP colliders,''
  Phys.\ Rev.\  D {\bf 54}, 5395 (1996)
  [arXiv:hep-ph/9605398].
  %%CITATION = PHRVA,D54,5395;%%
}

%\DineGU
\lref\DineGU{
  M.~Dine and W.~Fischler,
  ``A Phenomenological Model Of Particle Physics Based On Supersymmetry,''
  Phys.\ Lett.\  B {\bf 110}, 227 (1982).
  %%CITATION = PHLTA,B110,227;%%
}

%\NappiHM
\lref\NappiHM{
  C.~R.~Nappi and B.~A.~Ovrut,
  ``Supersymmetric Extension Of The SU(3) X SU(2) X U(1) Model,''
  Phys.\ Lett.\  B {\bf 113}, 175 (1982).
  %%CITATION = PHLTA,B113,175;%%
}

%\DineZB
\lref\DineZB{
  M.~Dine and W.~Fischler,
  ``A Supersymmetric Gut,''
  Nucl.\ Phys.\  B {\bf 204}, 346 (1982).
  %%CITATION = NUPHA,B204,346;%%
}

%\AlvarezGaumeWY
\lref\AlvarezGaumeWY{
  L.~Alvarez-Gaume, M.~Claudson and M.~B.~Wise,
  ``Low-Energy Supersymmetry,''
  Nucl.\ Phys.\  B {\bf 207}, 96 (1982).
  %%CITATION = NUPHA,B207,96;%%
}

%\DimopoulosGM
\lref\DimopoulosGM{
  S.~Dimopoulos and S.~Raby,
  ``Geometric Hierarchy,''
  Nucl.\ Phys.\  B {\bf 219}, 479 (1983).
  %%CITATION = NUPHA,B219,479;%%
}

%\DineZA
\lref\DineZA{
  M.~Dine, W.~Fischler and M.~Srednicki,
  ``Supersymmetric Technicolor,''
  Nucl.\ Phys.\  B {\bf 189}, 575 (1981).
  %%CITATION = NUPHA,B189,575;%%
}

%\DimopoulosAU
\lref\DimopoulosAU{
  S.~Dimopoulos and S.~Raby,
  ``Supercolor,''
  Nucl.\ Phys.\  B {\bf 192}, 353 (1981).
  %%CITATION = NUPHA,B192,353;%%
}

\lref\tobenomura{
  Y.~Nomura and K.~Tobe,
  ``Phenomenological aspects of a direct-transmission model of dynamical
  supersymmetry breaking with the gravitino mass m(3/2) < 1-keV,''
  Phys.\ Rev.\  D {\bf 58}, 055002 (1998)
  [arXiv:hep-ph/9708377].
}

\lref\IzawaGS{
  K.~I.~Izawa, Y.~Nomura, K.~Tobe and T.~Yanagida,
  ``Direct-transmission models of dynamical supersymmetry breaking,''
  Phys.\ Rev.\  D {\bf 56}, 2886 (1997)
  [arXiv:hep-ph/9705228].
  %%CITATION = PHRVA,D56,2886;%%
}

%\DimopoulosYQ
\lref\DimopoulosYQ{
  S.~Dimopoulos, S.~D.~Thomas and J.~D.~Wells,
  ``Sparticle spectroscopy and electroweak symmetry breaking with
  gauge-mediated supersymmetry breaking,''
  Nucl.\ Phys.\  B {\bf 488}, 39 (1997)
  [arXiv:hep-ph/9609434].
  %%CITATION = NUPHA,B488,39;%%
}

%\CulbertsonAM
\lref\CulbertsonAM{
  R.~Culbertson {\it et al.}  [SUSY Working Group Collaboration],
  ``Low-scale and gauge-mediated supersymmetry breaking at the Fermilab
  Tevatron Run II,''
  arXiv:hep-ph/0008070.
  %%CITATION = HEP-PH/0008070;%%
}

%\BaerPE
\lref\BaerPE{
  H.~Baer, P.~G.~Mercadante, X.~Tata and Y.~l.~Wang,
  ``The reach of the CERN Large Hadron Collider for gauge-mediated
  supersymmetry breaking models,''
  Phys.\ Rev.\  D {\bf 62}, 095007 (2000)
  [arXiv:hep-ph/0004001].
  %%CITATION = PHRVA,D62,095007;%%
}

%\BaerTX
\lref\BaerTX{
  H.~Baer, P.~G.~Mercadante, X.~Tata and Y.~l.~Wang,
  ``The reach of Tevatron upgrades in gauge-mediated supersymmetry breaking
  models,''
  Phys.\ Rev.\  D {\bf 60}, 055001 (1999)
  [arXiv:hep-ph/9903333].
  %%CITATION = PHRVA,D60,055001;%%
}

%\MatchevFT
\lref\MatchevFT{
  K.~T.~Matchev and S.~D.~Thomas,
  ``Higgs and Z-boson signatures of supersymmetry,''
  Phys.\ Rev.\  D {\bf 62}, 077702 (2000)
  [arXiv:hep-ph/9908482].
  %%CITATION = PHRVA,D62,077702;%%
}

%\NelsonNF
\lref\NelsonNF{
  A.~E.~Nelson and N.~Seiberg,
  ``R symmetry breaking versus supersymmetry breaking,''
  Nucl.\ Phys.\  B {\bf 416}, 46 (1994)
  [arXiv:hep-ph/9309299].
  %%CITATION = NUPHA,B416,46;%%
}

\lref\ISSi{
  K.~Intriligator, N.~Seiberg and D.~Shih,
  ``Dynamical SUSY breaking in meta-stable vacua,''
  JHEP {\bf 0604}, 021 (2006)
  [arXiv:hep-th/0602239].
  %%CITATION = JHEPA,0604,021;%%
}

%\GiudiceNI
\lref\GiudiceNI{
  G.~F.~Giudice and R.~Rattazzi,
  ``Extracting supersymmetry-breaking effects from wave-function
  renormalization,''
  Nucl.\ Phys.\  B {\bf 511}, 25 (1998)
  [arXiv:hep-ph/9706540].
  %%CITATION = NUPHA,B511,25;%%
}
\lref\agashe{
  K.~Agashe,
  ``Can multi-TeV (top and other) squarks be natural in gauge mediation?,''
  Phys.\ Rev.\  D {\bf 61}, 115006 (2000)
  [arXiv:hep-ph/9910497].
  %%CITATION = PHRVA,D61,115006;%%
}

\lref\agashegraesser{
  K.~Agashe and M.~Graesser,
  ``Improving the fine tuning in models of low energy gauge mediated
  supersymmetry breaking,''
  Nucl.\ Phys.\  B {\bf 507}, 3 (1997)
  [arXiv:hep-ph/9704206].
  %%CITATION = NUPHA,B507,3;%%
}

\lref\ournextpaper{
  C.~Cheung, A.~L.~Fitzpatrick, P.~Meade and D.~Shih, in
  preparation.
}

%\BarbieriFN
\lref\BarbieriFN{
  R.~Barbieri and G.~F.~Giudice,
  ``Upper Bounds On Supersymmetric Particle Masses,''
  Nucl.\ Phys.\  B {\bf 306}, 63 (1988).
  %%CITATION = NUPHA,B306,63;%%
}

%\KobayashiFH
\lref\KobayashiFH{
  T.~Kobayashi, H.~Terao and A.~Tsuchiya,
  ``Fine-tuning in gauge mediated supersymmetry breaking models and induced
  top Yukawa coupling,''
  Phys.\ Rev.\  D {\bf 74}, 015002 (2006)
  [arXiv:hep-ph/0604091].
  %%CITATION = PHRVA,D74,015002;%%
}

\lref\AitchisonCF{
  I.~J.~R.~Aitchison,
  ``Supersymmetry and the MSSM: An elementary introduction,''
  arXiv:hep-ph/0505105.
  %%CITATION = HEP-PH/0505105;%%
}

\lref\PeskinEZ{
  M.~E.~Peskin,
  ``Beyond the standard model,''
  arXiv:hep-ph/9705479.
  %%CITATION = HEP-PH/9705479;%%
}

%\ORaifeartaighPR
\lref\ORaifeartaighPR{
  L.~O'Raifeartaigh,
  ``Spontaneous Symmetry Breaking For Chiral Scalar Superfields,''
  Nucl.\ Phys.\  B {\bf 96}, 331 (1975).
  %%CITATION = NUPHA,B96,331;%%
}

\lref\KitanoXG{
  R.~Kitano, H.~Ooguri and Y.~Ookouchi,
  ``Direct mediation of meta-stable supersymmetry breaking,''
  Phys.\ Rev.\  D {\bf 75}, 045022 (2007)
  [arXiv:hep-ph/0612139].
  %%CITATION = PHRVA,D75,045022;%%
}

\lref\CsakiWI{
  C.~Csaki, Y.~Shirman and J.~Terning,
  ``A simple model of low-scale direct gauge mediation,''
  JHEP {\bf 0705}, 099 (2007)
  [arXiv:hep-ph/0612241].
  %%CITATION = JHEPA,0705,099;%%
}

\lref\AmaritiQU{
  A.~Amariti, L.~Girardello and A.~Mariotti,
  ``On meta-stable SQCD with adjoint matter and gauge mediation,''
  Fortsch.\ Phys.\  {\bf 55}, 627 (2007)
  [arXiv:hep-th/0701121].
  %%CITATION = FPYKA,55,627;%%
}

\lref\MurayamaYF{
  H.~Murayama and Y.~Nomura,
  ``Gauge mediation simplified,''
  Phys.\ Rev.\ Lett.\  {\bf 98}, 151803 (2007)
  [arXiv:hep-ph/0612186].
  %%CITATION = PRLTA,98,151803;%%
}

\lref\KitanoWM{
  R.~Kitano,
  ``Dynamical GUT breaking and mu-term driven supersymmetry breaking,''
  Phys.\ Rev.\  D {\bf 74}, 115002 (2006)
  [arXiv:hep-ph/0606129].
  %%CITATION = PHRVA,D74,115002;%%
}

\lref\KitanoWZ{
  R.~Kitano,
  ``Gravitational gauge mediation,''
  Phys.\ Lett.\  B {\bf 641}, 203 (2006)
  [arXiv:hep-ph/0607090].
  %%CITATION = PHLTA,B641,203;%%
}

\lref\DineXT{
  M.~Dine and J.~Mason,
  ``Gauge mediation in metastable vacua,''
  arXiv:hep-ph/0611312.
  %%CITATION = HEP-PH/0611312;%%
}

\lref\AharonyMY{
  O.~Aharony and N.~Seiberg,
  ``Naturalized and simplified gauge mediation,''
  JHEP {\bf 0702}, 054 (2007)
  [arXiv:hep-ph/0612308].
  %%CITATION = JHEPA,0702,054;%%
}

\lref\alephref{
  A.~Heister {\it et al.}  [ALEPH Collaboration],
  ``Single- and multi-photon production in e+ e- collisions at s**(1/2) up to
  209-GeV,''
  Eur.\ Phys.\ J.\  C {\bf 28}, 1 (2003).
  %%CITATION = EPHJA,C28,1;%%
}

%\ShihAV
\lref\ShihAV{
  D.~Shih,
  ``Spontaneous R-symmetry breaking in O'Raifeartaigh models,''
  arXiv:hep-th/0703196.
  %%CITATION = HEP-TH/0703196;%%
}

\lref\pdgRaby{
    S.~Raby,
  ``Grand Unified Theories,'', {\it in}
  W.~M.~Yao {\it et al.}  [Particle Data Group],
  ``Review of particle physics,''
  J.\ Phys.\ G {\bf 33}, 1 (2006).
}

\lref\martinvaughn{
   S.~P.~Martin and M.~T.~Vaughn,
  ``Two Loop Renormalization Group Equations For Soft Supersymmetry Breaking
  Couplings,''
  Phys.\ Rev.\  D {\bf 50}, 2282 (1994)
  [arXiv:hep-ph/9311340].
}
\lref\matchev{
  D.~M.~Pierce, J.~A.~Bagger, K.~T.~Matchev and R.~j.~Zhang,
  ``Precision corrections in the minimal supersymmetric standard model,''
  Nucl.\ Phys.\  B {\bf 491}, 3 (1997)
  [arXiv:hep-ph/9606211].
}

\lref\AllanachKG{
  B.~C.~Allanach,
  ``SOFTSUSY: A C++ program for calculating supersymmetric spectra,''
  Comput.\ Phys.\ Commun.\  {\bf 143}, 305 (2002)
  [arXiv:hep-ph/0104145].
  %%CITATION = CPHCB,143,305;%%
}

%\PoppitzVD
\lref\PoppitzVD{
  E.~Poppitz and S.~P.~Trivedi,
  ``Dynamical supersymmetry breaking,''
  Ann.\ Rev.\ Nucl.\ Part.\ Sci.\  {\bf 48}, 307 (1998)
  [arXiv:hep-th/9803107].
  %%CITATION = ARNUA,48,307;%%
}

%\TerningTH
\lref\TerningTH{
  J.~Terning,
  ``Non-perturbative supersymmetry,''
  arXiv:hep-th/0306119.
  %%CITATION = HEP-TH/0306119;%%
}

%\MartinZB
\lref\MartinZB{
  S.~P.~Martin,
  ``Generalized messengers of supersymmetry breaking and the sparticle mass
  spectrum,''
  Phys.\ Rev.\  D {\bf 55}, 3177 (1997)
  [arXiv:hep-ph/9608224].
  %%CITATION = PHRVA,D55,3177;%%
}

%\ShadmiJY
\lref\ShadmiJY{
  Y.~Shadmi and Y.~Shirman,
  ``Dynamical supersymmetry breaking,''
  Rev.\ Mod.\ Phys.\  {\bf 72}, 25 (2000)
  [arXiv:hep-th/9907225].
  %%CITATION = RMPHA,72,25;%%
}

%\AffleckUZ
\lref\AffleckUZ{
  I.~Affleck, M.~Dine and N.~Seiberg,
  ``Calculable Nonperturbative Supersymmetry Breaking,''
  Phys.\ Rev.\ Lett.\  {\bf 52}, 1677 (1984).
  %%CITATION = PRLTA,52,1677;%%
}
%\AffleckXZ
\lref\AffleckXZ{
  I.~Affleck, M.~Dine and N.~Seiberg,
  ``Dynamical Supersymmetry Breaking In Four-Dimensions And Its
  Phenomenological Implications,''
  Nucl.\ Phys.\ B {\bf 256}, 557 (1985).
  %%CITATION = NUPHA,B256,557;%%
}

%\AbelJX
\lref\AbelJX{
  S.~Abel, C.~Durnford, J.~Jaeckel and V.~V.~Khoze,
  ``Dynamical breaking of $U(1)_{R}$ and supersymmetry in a metastable vacuum,''
  arXiv:0707.2958 [hep-ph].
  %%CITATION = ARXIV:0707.2958;%%
}

%\HabaRJ
\lref\HabaRJ{
  N.~Haba and N.~Maru,
  ``A Simple Model of Direct Gauge Mediation of Metastable Supersymmetry
  Breaking,''
  arXiv:0709.2945 [hep-ph].
  %%CITATION = ARXIV:0709.2945;%%
}

%\FerrettiRQ
\lref\FerrettiRQ{
  L.~Ferretti,
  ``O'Raifeartaigh models with spontaneous R-symmetry breaking,''
  arXiv:0710.2535 [hep-th].
  %%CITATION = ARXIV:0710.2535;%%
}

%\DvaliCU
\lref\DvaliCU{
  G.~R.~Dvali, G.~F.~Giudice and A.~Pomarol,
  ``The $\mu$-Problem in Theories with Gauge-Mediated Supersymmetry Breaking,''
  Nucl.\ Phys.\  B {\bf 478}, 31 (1996)
  [arXiv:hep-ph/9603238].
  %%CITATION = NUPHA,B478,31;%%
}

%\DimopoulosIG
\lref\DimopoulosIG{
  S.~Dimopoulos and G.~F.~Giudice,
  ``Multi-messenger theories of gauge-mediated supersymmetry breaking,''
  Phys.\ Lett.\  B {\bf 393}, 72 (1997)
  [arXiv:hep-ph/9609344].
  %%CITATION = PHLTA,B393,72;%%
}

%%%%%%%%%%%%%%%%%%%%%%%%

\newbox\tmpbox\setbox\tmpbox\hbox{\abstractfont }
\Title{\vbox{\baselineskip12pt }} {\vbox{\centerline{
(Extra)Ordinary Gauge Mediation
}}}
\smallskip
\centerline{Clifford Cheung,$^{1,2}$ A. Liam Fitzpatrick,$^1$ and
David Shih$^2$}
\smallskip
\bigskip
\centerline{{\it $^1$Department of Physics, Harvard University,
Cambridge, MA 02138 USA}}
\smallskip
\centerline{{\it $^2$School of Natural Sciences, Institute for
Advanced Study, Princeton, NJ 08540 USA}}
\bigskip
\vskip 1cm

\noindent We study models of ``(extra)ordinary gauge mediation,"
which consist of taking ordinary gauge mediation and extending the
messenger superpotential to include all renormalizable couplings
consistent with SM gauge invariance and an R-symmetry. We classify
all such models and find that their phenomenology can differ
significantly from that of ordinary gauge mediation. Some
highlights include: arbitrary modifications of the squark/slepton
mass relations, small $\mu$ and Higgsino NLSP's, and the
possibility of having fewer than one effective messenger. We also
show how these models lead naturally to extremely simple examples
of direct gauge mediation, where SUSY and R-symmetry breaking
occur not in a hidden sector, but due to the dynamics of the
messenger sector itself.

\bigskip

\Date{October 2007}

\newsec{Introduction}

\subsec{Motivation}

The LHC is coming, and the question on everyone's mind is: what
will we see? One reasonable guess is supersymmetry, probably still
the most compelling candidate for physics beyond the standard
model. The minimal incarnation of SUSY is the MSSM, but this is
only an incomplete phenomenological framework. (For a nice review
of the MSSM, see e.g.\ \martinreview.) Soft SUSY breaking in the
MSSM introduces $\sim 100$ new couplings in addition to those of
the standard model, and in their most generic form, these new
couplings give rise to serious flavor and CP problems. Thus, even
if we discover the MSSM at the LHC, we will still have the main
theoretical challenge ahead of us: explaining the origin of the
MSSM parameters with an underlying model of SUSY breaking that is
consistent with flavor and CP.

Gauge mediation \refs{\DineZA\DimopoulosAU\DineGU\NappiHM\DineZB\AlvarezGaumeWY\DimopoulosGM\DineYW\DineVC-\DineAG}
(see also \refs{\GiudiceBP\PoppitzVD\TerningTH-\ShadmiJY} for reviews, and
many relevant references) is a particularly attractive way of
generating soft SUSY breaking in the MSSM. Not only does it solve
the flavor and CP problems, but it is also calculable, predictive,
and phenomenologically distinctive. Over the years, a great deal
of work has been devoted to building complete models of gauge
mediation, spurred by theoretical progress in constructing
calculable examples \refs{\AffleckUZ, \AffleckXZ} of dynamical
SUSY breaking \WittenNF. As a result, there are now many viable
models of gauge mediation, complete with detailed hidden sectors
where SUSY is broken dynamically through strong gauge dynamics.

The study of the low-energy phenomenology of gauge mediation has
proceeded in conjunction with these model-building efforts. Since
the details of the hidden sector are often phenomenologically
irrelevant,\foot{This is not always the case, as was recently
pointed out in \CohenQC.} people here have mostly relied on a
simplified, incomplete framework known as ``ordinary gauge
mediation'' (OGM), where the hidden sector is parameterized by a
singlet field $X$ which is a spurion for SUSY breaking,
\eqn\Xvev{ \langle X\rangle=X+\theta^2F, } (We will use $X$ to
denote both the superfield and the vev of its lowest component.)
OGM also includes $N$ vector-like pairs of messenger fields
$\phi_i$, $\tilde\phi_i$, transforming in the ${\bf 5}\oplus {\bf
\bar 5}$ representations under $SU(5)\supset G_{\rm
SM}$.\foot{This is the simplest matter content consistent with
gauge coupling unification. Other representations are also
possible, including those that do not come in complete GUT
multiplets \MartinZB, but we will not consider these here.} The
messengers interact with $X$ via Yukawa-like couplings
\eqn\OGMintro{\eqalign{
  W &= \lambda_{ij} X \phi_i\tilde\phi_j
}}
where the sum on $i$, $j=1,\,\dots,\, N$ is implicit. (Gauge
indices are suppressed here and throughout.)
Through \OGMintro\ and the gauge interactions, the messengers
communicate SUSY breaking from the hidden sector to the MSSM. The
result is an MSSM spectrum with many distinctive features, some of
which we will review later in this introduction.

Given that much of the classic low-energy phenomenology of gauge
mediation has been derived using the framework of OGM, it is
important to ask (especially in the LHC era): is OGM truly
representative of gauge mediation in general, or is it only one of
many possible gauge mediation phenomenologies? In particular, how
do things change if we deform or extend OGM in various directions?

In this paper, we would like to address these questions by
studying a large family of extensions of OGM, obtained by generalizing \OGMintro\ to include all renormalizable, gauge invariant couplings between the messengers and any number of singlet fields.
Since these models extend OGM into a wider parameter space, yet they are no less ``ordinary"
by any sensible measure (i.e.\ they are renormalizable and are not forbidden by any symmetries or experimental constraints),
we will refer to them as models of ``(extra)ordinary gauge mediation" (EOGM).
In the following sections, we will present an in-depth study of the phenomenology of EOGM, and we will see that
it can differ in interesting ways from that of OGM.

\subsec{The phenomenology of (extra)ordinary gauge mediation}

Now let us describe our EOGM models and their phenomenology in
more detail. We start with the most general renormalizable, gauge
invariant superpotential describing the couplings between the
messengers and any number of singlets $X_k$: \eqn\EOGMmultising{ W
= (\lambda_{ij}^{(k)}X_k + M_{ij})\phi_i\tilde\phi_j =
(\lambda_{2ij}^{(k)} X_k + M_{2ij}) \ell_i \tilde{\ell}_j +
(\lambda_{3ij}^{(k)} X_k + M_{3ij}) q_i \tilde{q}_j } where in the
second equation of \EOGMmultising, we have decomposed $\phi_i$,
$\tilde\phi_i$ into their $SU(2)$ doublet and $SU(3)$ triplet
components, $\ell_i$, $\tilde \ell_i$ and $q_i$, $\tilde q_i$,
respectively. We emphasize that doublet/triplet splitting in
\EOGMmultising\ is similar in spirit to the doublet/triplet
splitting that already happens in SUSY GUT embeddings of the MSSM
(indeed they may very well have the same origin), so there is
really no reason not to consider the most general form of
\EOGMmultising.

In fact, this model can be reduced to a model with only one singlet, through the following trivial observation. Through a unitary transformation, we can always rotate the singlet fields so that only one of them, call it $X$, acquires a SUSY-breaking F-component vev as in \Xvev. Then the remaining singlets only have scalar component vevs, $\langle X_k\rangle = X_k$, and since we are only interested in the tree-level messenger mass matrix, we are free to substitute these into the superpotential \EOGMmultising. This reduces it to the form
\eqn\EOGMintro{\eqalign{
 W &= (\lambda_{ij} X+m_{ij}) \phi_i\tilde\phi_j
  = (\lambda_{2ij} X + m_{2ij}) \ell_i \tilde{\ell}_j + (\lambda_{3ij} X + m_{3ij}) q_i \tilde{q}_j
}}
In other words, we have shown that the most general EOGM model is simply OGM plus arbitrary supersymmetric mass terms for the messengers.

Surprisingly, while there are many examples in the
literature of OGM deformed by mass terms (including many of the original models of gauge mediation
\refs{\DineGU\NappiHM\DineZB\AlvarezGaumeWY-\DimopoulosGM}, some more modern
models \refs{\tobenomura\IzawaGS\DvaliCU\DimopoulosIG-\DimopoulosYQ }, and most recently many of the models
based on \ISSi),
the phenomenology of these models has not been explored in any systematic
way.\foot{Perhaps one reason for this is that generic multi-messenger models are problematic, as they generally have tachyonic one-loop slepton masses coming from contractions of the hypercharge D-terms. (We thank M.~Dine for pointing out this effect to us.) We will discuss this problem -- and how we get around it --  in more detail in section 2.} In this paper, we will take the first steps in this
direction.

To simplify our analysis, and because it has some distinctive and desirable consequences,
we will limit our study to models possessing a non-trivial $U(1)_R$ symmetry, which is only broken spontaneously by the vev of $X$ \Xvev.\foot{
These models always possess a trivial R-symmetry under which $R(X)=0$
and $R(\phi_i)=R(\tilde\phi_i)=1$. The $U(1)_R$ we are imposing on \EOGMintro\ is in addition to this, and
it results in various selection rules on the couplings $m_{ij}$, $\lambda_{ij}$.}
We will show that in a general EOGM model with an R-symmetry, the soft
masses at the messenger scale are given by a simple generalization
of the usual OGM formulae
\eqn\EOGMmasses{\eqalign{
 & M_r = {\alpha_r\over 4\pi}\Lambda_G,\qquad m_{\tilde f}^2 = 2\sum_{r=1}^3 C_{\tilde f}\,^r
\left({\alpha_r\over
 4\pi}\right)^2 {\Lambda_{G}^2} N_{{\rm eff},r}^{-1}
 }}
Here $\Lambda_G\sim F/X$ sets the overall scale of the soft
masses, $N_{{\rm eff},2}$
and $N_{{\rm eff},3}$ can be thought of as ``effective" doublet
and triplet messenger numbers, and $N_{{\rm eff},1}^{-1} \equiv {3\over 5}N_{{\rm
eff},2}^{-1}+{2\over 5}N_{{\rm eff},3}^{-1}$.\foot{The rest of the notation is as in
\GiudiceBP. In particular, $r=1$, 2, 3 labels the SM gauge groups
$U(1)$, $SU(2)$ and $SU(3)$, respectively; $\tilde f$ labels an
MSSM sfermion field; and $C_{\tilde f}\,^r$ is the quadratic
Casimir of $\tilde f$ in the gauge group $r$.} In general, $N_{{\rm eff},2}$ and $N_{{\rm eff},3}$  depend on all the
doublet and triplet parameters of the model, respectively:
\eqn\Neffintro{
N_{{\rm eff},r}\equiv N_{{\rm eff}}(X,m_{r},\lambda_{r}) \qquad
(r=2,\,3)
}
and they take values between 0 and $N$ inclusive. (The full
formula for $N_{\rm eff}$ can be found in section 2.) Note that
the special case of OGM corresponds to $N_{{\rm eff},2}=N_{{\rm
eff},3}=N$ -- the messenger numbers in this case are equal and are
independent of all the couplings.

The effective messenger numbers play an important role in
determining the low-energy phenomenology of EOGM. In particular,
\Neffintro\ implies that doublet/triplet splitting can lead to
different effective messenger numbers for doublets and triplets
(unlike in OGM), and this in turn can have a large, qualitative
effect on the spectrum. Some specific ways in which EOGM can
deviate from OGM include:

\lfm{1.} {\it Modified relations between squark and slepton
masses.} Typically, in gauge mediation, the squark mass-squareds
are always much larger than the slepton mass-squareds, since
$\alpha_3\gg \alpha_2$, $\alpha_1$. However, by making $N_{\rm
eff,3}\gg N_{\rm eff,2}$, the sfermion masses can be squashed
together, as can be seen from \EOGMmasses.

\lfm{2.} {\it The possibility for small $\mu$ and Higgsino NLSPs
in a large portion of parameter space.} A more subtle consequence
of having different doublet and triplet messenger numbers is that
this can lead to small $\mu$ through a cancellation in the running
of $m_{H_u}^2$ \refs{\agashe,\agashegraesser}. Aside from its
possible implications for the little hierarchy problem, small
$\mu$ in gauge mediation is interesting because it implies that
the NLSP is a Higgsino-like neutralino. This novel scenario has
not been studied much in the past (see however
\refs{\DimopoulosYQ,\BaerTX\MatchevFT\BaerPE-\CulbertsonAM}),
presumably because in OGM the NLSP is always either the bino or
the stau.

\lfm{3.} {\it Effective messenger number less than one.} In the
space of EOGM models, one can achieve $N_{\rm eff}<1$, which is
obviously never possible in OGM where $N_{\rm eff}=N$. This is
interesting, as it allows the gauginos to be lighter than in any
OGM scenario. Lighter gluinos, in particular, could significantly
enhance sparticle production at the LHC, relative to standard OGM rates.

\lfm{4.} {\it Gauge coupling unification.} We will see that in
these models, the R-symmetry allows for gauge coupling unification
to be maintained without tuning of parameters, even with different
effective numbers of doublet and triplet messengers. This is a
crucial difference between these models and those of
\refs{\agashe,\agashegraesser}, where additional doublet and/or
triplet fields were put in by hand to ensure unification.

\bigskip

Finally, let us mention one aspect of the spectrum that does not
change between OGM and EOGM models (with an R-symmetry). According
to \EOGMmasses, the gaugino masses always obey the GUT relations
in these models, $M_1:M_2:M_3=\alpha_1:\alpha_2:\alpha_3$,
regardless of the amount of doublet/triplet splitting. As we will
see in the next section, this is a direct consequence of imposing
a non-trivial R-symmetry on the superpotential \EOGMintro\ under which $R(X)\ne 0$. In more general
models without such an R-symmetry, even the gaugino mass relations can
be modified arbitrarily through doublet/triplet splitting.

\subsec{Minimal completions of gauge mediation}

In addition to exploring the phenomenology of gauge mediation,
there is another, more formal motivation for studying models of
the form \EOGMintro: the goal of finding simple examples of
``direct gauge mediation," i.e. models in which the messengers are
also part of the SUSY breaking sector. Indeed, our EOGM models can
be trivially completed into generalized O'Raifeartaigh models of
the kind discussed recently in \ShihAV, simply by adding $\delta
W=F X$ to \EOGMintro:
\eqn\EOGMintroiii{
 W = \lambda_{ij}X\phi_i\tilde\phi_j +
 m_{ij}\phi_i\tilde\phi_j+F X
 }
As we will see, the R-symmetry guarantees that the tree-level
scalar potential has a pseudo-moduli space of SUSY-breaking local
minima, located at $\phi=\tilde\phi=0$ and $|X|$ in some window.
At one-loop, a Coleman-Weinberg potential is generated on the
pseudo-moduli space, and the minima of this potential (if they
exist) are SUSY-breaking vacua of the theory.

In order for these models to be phenomenologically viable, the
R-symmetry must be spontaneously broken in the vacuum (otherwise
the gauginos cannot obtain soft masses). We will see that such
R-symmetry breaking minima of the CW potential can exist in the
parameter space of these models, because there are typically
fields with R-charge $R\ne 0$, 2 \ShihAV. Therefore, these models
can serve as extremely compact examples of direct gauge mediation,
which are complete in the sense that the sources of SUSY and
R-symmetry breaking are included. Note that these models are {\it
not} examples of dynamical SUSY breaking, nor do they explain the
origin of $\mu$ and $B\mu$. However, they do provide a minimal
framework in which these issues can be further explored.

\subsec{Outline}

The outline of our paper is as follows. In section 2 we will
discuss some general aspects of EOGM, including: formulae for
$N_{\rm eff}(X,m,\lambda)$ and the MSSM soft masses; a discussion
of doublet/triplet splitting and its effects; and the issue of
gauge coupling unification. In section 3 we will introduce a
classification of EOGM models. We will see that the models fall
into three distinct categories which have qualitatively different
phenomenology. In section 4 we will analyze in detail the
phenomenology of some simple examples of EOGM models and show how
some of the general features discussed in section 2 can be
realized. Finally, section 5 contains an analysis of the minimal
completions \EOGMintroiii.

In appendix A, we prove some useful results about the mass matrix
of the messengers, which have implications for the MSSM soft
SUSY-breaking terms. Appendix B has a discussion of our treatment
of the MSSM RGEs, a careful understanding of which is important
for obtaining accurate low-energy MSSM spectra. In appendix C,
there are some useful formulae for the neutralino and chargino
mass matrices in the small $\mu$ limit, as well as a very
preliminary discussion of the collider phenomenology of Higgsino
NLSPs.

\newsec{General Aspects of (Extra)Ordinary Gauge Mediation}

\subsec{The models}

In this section, we would like to study general aspects of the
phenomenology of EOGM models. As discussed in the introduction,
the models consist of a singlet $X$ and $N$ messengers $\phi_i$,
$\tilde\phi_i$ transforming in the ${\bf 5}\oplus {\bf \bar 5}$
representation of $SU(5)\supset G_{\rm SM}$. Through some
unspecified dynamics in the hidden sector, $X$ acquires a SUSY-
and R-symmetry-breaking vev, $\langle X\rangle = X+\theta^2 F$.
The couplings between $X$ and the messengers are described by the
most general superpotential consistent with renormalizability, SM
gauge invariance, and a non-trivial R-symmetry:
 \eqn\WORii{ W= \CM_{ij}(X) \phi_i\tilde\phi_j =
 (\lambda_{ij}X+m_{ij})\phi_i\tilde\phi_j
 }
where $\CM_{ij}(X)= \lambda_{ij} X+m_{ij}$ is the messenger mass
matrix, and the R-symmetry means that the couplings in this mass matrix
must obey a set of selection rules.\foot{Note that although we are imposing this
R-symmetry on the messenger superpotential, it could actually be
an accidental symmetry of the underlying, strongly-coupled gauge
theory which presumably dynamically generates all the mass scales
in \EOGMintro\ (and in which $X$ and/or the messengers could be
composite fields). This is precisely what happens, for instance,
in massive SQCD in the free-magnetic phase \ISSi.} Let us now describe these selection rules in more detail.
For the time being, we will assume for simplicity
that the couplings in \WORii\ respect the full $SU(5)$ invariance;
in section 2.3 and beyond, we will consider the effect of
doublet/triplet splitting in detail.

Suppose that \WORii\ respects a non-trivial R-symmetry under which the fields transform
with R-charges $R(X)\ne 0$, $R(\phi_i)$, $R(\tilde\phi_i)$. (This will be the case for all the models studied in this
paper.) Then the selection rules take the form
\eqn\Rselintro{\eqalign{
 & \lambda_{ij}\ne 0\quad {\rm only\,\,\,if}\quad
R(\phi_i)+R(\tilde\phi_j)=2-R(X) \cr
 & m_{ij}\ne 0\quad {\rm
only\,\,\,if}\quad R(\phi_i)+R(\tilde\phi_j)=2
 }}
since $W$ must always have definite R-charge $R(W)=2$,

These selection rules, and the R-symmetry more generally, have
many important consequences which we will explore in the following
subsections, starting with the spectrum of MSSM soft masses. Most
of these consequences stem from a non-trivial identity satisfied
by the messenger mass matrix,
 \eqn\detmmess{
  \det\, \CM = X^n G(m,\lambda),\qquad n = {1\over R(X)} \sum_{i=1}^N(2-R(\phi_i)-R(\tilde \phi_i)),
 }
where $G(m,\lambda)$ is some function of the couplings. This
identity follows directly from the selection rules \Rselintro; for
a straightforward proof, see appendix A. Note that in this
identity, $n$ must be an integer satisfying $0 \le n \le N$, since
$\det(\lambda X+m)$ is a degree $N$ polynomial in $X$.

Although we have allowed $R(X)$ to take any non-zero value in the discussion of the R-symmetry so far, it turns out that not all of these R-symmetries are distinct. In fact, if the model is invariant under
an R-symmetry with $R(X)\ne 0$, then it must be
invariant under a continuous family of {\it equivalent} R-symmetries parametrized by arbitrary $R(X)\in {\Bbb R}$.
These are obtained by mixing the R-symmetry with
the trivial $U(1)_R$ that is always respected by \WORii, under which $R(X)=0$, $R(\phi_i)=R(\tilde\phi_i)=1$. (As a consistency
check, note that
the formula for $n$ in \detmmess\ remains invariant under this mixing of R-symmetries.)
In particular, we can always use this to set
\eqn\RofX{
R(X)=2
}
without loss of generality. Henceforth, we will assume this implicitly in the paper. This will turn out to be a convenient choice in section 5,
where we analyze the ``complete" models obtained by perturbing \WORii\ by the SUSY-breaking linear term $\delta W = F X$.

\subsec{MSSM soft masses}

It is straightforward to derive formulae for the running gaugino
and sfermion soft masses at the messenger scale, by generalizing
the wavefunction renormalization technique of \GiudiceNI. For the
gaugino masses we find (using the determinant identity \detmmess)
\eqn\mgauginogen{\eqalign{
 & M_r = {\alpha_r \over 4\pi}
 \Lambda_G,\qquad \Lambda_G = F\,\partial_X \log\det\, \CM =
{n F\over
 X}
}}
while the sfermion masses are given by
\eqn\scalmassgen{\eqalign{
 m_{\tilde f}^2 = 2 \sum_{r=1}^3 C_{\tilde
f}\,^r\left({\alpha_r\over 4\pi}\right)^2 \Lambda_S^2,\qquad
 \Lambda_S^2 = \half |F|^2\,{\partial^2\over\partial X\partial
X^*}\sum_{i=1}^{N} \left(\log |\CM_i|^2\right)^2
 }}
where $\CM_i$ denote the eigenvalues of $\CM$. (The rest of the
notation is described in the introduction.) In these formulas, the
gauge couplings $\alpha_r$ are all evaluated at the messenger
scale. In order to find the physical spectrum, one must of course
run everything down to the weak scale. Our procedure for this is
described in appendix B.

The soft masses \mgauginogen\ and \scalmassgen\ are
generalizations of well-known OGM formulae. (See e.g.\ \GiudiceBP, whose
conventions we largely adhere to in this paper.) By analogy with OGM,
it is useful to define the ``effective messenger number'' to be
 \eqn\Neffdef{ N_{\rm eff}(X,m,\lambda)
\equiv {\Lambda_G^2\over\Lambda_S^2} = \left[{1\over 2n^2}
|X|^2{\partial^2\over\partial X\partial X^*}\sum_{i=1}^{N}
\left(\log {|\CM_i|^2\over\mu^2}\right)^2\right]^{-1}
 }
In OGM, $N_{\rm eff}=N$, but more generally it is a continuous
function of the couplings taking values between 0 and $N$
inclusive.

A fact that will be useful in later sections is that $N_{\rm eff}$
simplifies somewhat in the asymptotic limits $X\rightarrow 0$,
$\infty$. In appendix A, we derive formulas for $N_{\rm eff}$ in
these limits. Here let us simply highlight two features of these
formulas that we will need later. First of all, the asymptotic
values of $N_{\rm eff}$ are independent of all the parameters,
\eqn\Neffasymconst{
\lim_{X\to 0,\,\infty}N_{\rm eff}(X,m,\lambda)=const.
}
Second, the asymptotic values of $N_{\rm eff}$ satisfy the
inequalities
\eqn\Neffasymineqi{\eqalign{
{n^2\over n^2-(N-r_m-1)(2n-N+r_m)}\,\le\, N_{\rm eff}(X\to
0)\,\le\, N-r_m
}}
and
\eqn\Neffasymineqii{\eqalign{
 {n^2\over r_\lambda+(r_\lambda-n)^2}\,\le \,N_{\rm eff}(X\to\infty)\,\le\, {n^2\over
 r_\lambda+{(r_\lambda-n)^2\over (N-r_\lambda)}}
 }}
where we have introduced the notation
\eqn\ranksdef{
r_\lambda\equiv {\rm rank}\,\lambda,\qquad r_m\equiv{\rm rank}\,m
}
This notation will also prove to be useful below.\foot{Note that
when $r_\lambda=N$, the upper bound on $N_{\rm eff}(X \to \infty)$
in \Neffasymineqii\ no longer makes sense. However, as we will
discuss more fully in section 3.2 below, in this case one always
has $N_{\rm eff}(X\to \infty)=N$.}

Finally, let us conclude this subsection by pointing out two effects that we have
ignored in writing down our formulae  \mgauginogen, \scalmassgen\ for the MSSM soft masses. The first is the effect of multiple messenger scales.
These can modify the formulae for the soft masses through
RG evolution, but in general this is a small effect. Below, in our
more quantitative analysis of specific examples, we will fully
account for the multiple messenger thresholds.
The second effect we are ignoring is the contribution to $\Lambda_G$, $\Lambda_S$
from higher-order corrections in $F/M_{\rm mess}^2$, where $M_{\rm
mess}$ is the (lightest) messenger scale. These cannot be
extracted from wavefunction renormalization, but instead require a
full Feynman diagram calculation. In the following we will assume
implicitly that $F\ll M_{\rm mess}^2$, in which case these
corrections are negligible.

\subsec{Doublet/triplet splitting and the MSSM soft masses}

So far, we have assumed for simplicity that the couplings in the
superpotential \WORii\ respect the full $SU(5)$ gauge symmetry.
However, the most general superpotential need only respect the SM
gauge symmetry; thus we are led to consider
\eqn\splitW{\eqalign{
W &=(\lambda_{2ij} X+m_{2ij}) \ell_i \tilde{\ell}_j+ (
\lambda_{3ij} X+m_{3ij} ) q_i \tilde{q}_j\cr
 }}
where $\ell$, $\tilde \ell$ and $q$, $\tilde q$ denote $SU(2)$
doublets and $SU(3)$ triplets, respectively. In this subsection,
we would like to describe the effect of doublet/triplet splitting
on the MSSM soft masses. Throughout, we will assume that the
doublet and triplet messengers have the same R-charge assignments.
As a result, the doublet and triplet messenger mass matrices will
always have the same structure and will both obey \detmmess\ with
the same $n$ and the same function $G$.

As discussed in the introduction, doublet/triplet
splitting has little effect on the MSSM soft masses in OGM. Thus, even
allowing for arbitrary doublet/triplet splitting, OGM leads to
very distinctive relations among the gaugino and the sfermion
masses. In EOGM, the relations amongst the gaugino masses are
still preserved,
\eqn\GUTrelmg{
M_1:M_2:M_3=\alpha_1:\alpha_2:\alpha_3
}
even with an arbitrary amount of doublet/triplet splitting. This
follows from \mgauginogen, according to which $\Lambda_G$ is
independent of the couplings and only depends on the integer $n$
(which is the same between the doublets and triplets). Let us
emphasize that this is a direct consequence of imposing on the model \WORii\ a non-trivial
R-symmetry under which $R(X)\ne 0$; if we abandon this symmetry,
then the GUT relations for the gaugino masses need no longer hold.\foot{M.\ Dine has pointed out to us that
in EOGM models which do not have an R-symmetry, doublet/triplet splitting can in general lead to
dangerous $\CO(1)$ CP-violating phases in the gaugino masses. Thus, avoiding these phases could be viewed as
another motivation for imposing a non-trivial R-symmetry on the space of EOGM models.}

 Next let us consider the sfermion masses. Here we have, instead
 of \scalmassgen:
 \eqn\scalmassdt{\eqalign{
 & m_{\tilde f}^2 = 2 \sum_{r=1}^3 C_{\tilde
f}\,^r\left({\alpha_r\over 4\pi}\right)^2 \Lambda_{Sr}^2
}}
with
\eqn\LambdaSdt{\eqalign{
 &  \Lambda_{S2}^2= \Lambda_{G}^2 N_{\rm eff}(X,m_2,\lambda_2)^{-1},\quad
  \Lambda_{S3}^2= \Lambda_{G}^2 N_{\rm eff}(X,m_3,\lambda_3)^{-1}
 }}
and
$\Lambda_{S1}^2={2\over5}\Lambda_{S3}^2+{3\over5}\Lambda_{S2}^2$.
Thus, the mass relations amongst the sfermions can be arbitrarily
modified through doublet/triplet splitting. In particular, by
taking $N_{{\rm eff},3}\gg N_{{\rm eff},2}$, the squark and
slepton masses can be brought closer together than in OGM (where
typically $m_{\tilde t}/m_{\tilde e_R}\sim 7$--10). This could be
helpful for solving the ``little hierarchy problem" of OGM, where
-- independent of the LEP bound on the Higgs mass -- the squarks
must be at least 700 GeV given the experimental lower bound of
$\sim 100$ GeV on the selectron mass.

Note that in writing down \scalmassdt, we have not included the dangerous contributions to the sfermion masses coming from contractions of the messenger hypercharge D-terms $D= g'(\phi^\dagger Y_\phi \phi - \tilde\phi^T Y_{\phi}\tilde\phi^*)$ \DineGU. 
If present, these contributions cause either the right or the left-handed sleptons to become tachyonic, because they are not positive definite, and they appear already at one-loop in the gauge interactions. They are absent in OGM, but unfortunately not in a generic EOGM model with arbitrary doublet/triplet splitting. In order to forbid these D-term contributions, we will impose on all the models we study in this paper the ``messenger parity" symmetry proposed in \DimopoulosIG
\eqn\messpar{
\phi\to U^*\tilde\phi^*,\qquad \tilde\phi\to\tilde U\phi^*,\qquad V\to -V
}
where $V$ stands for the SM gauge superfields, and $U$ and $\tilde U$ are some $N\times N$ unitary matrices. This is a symmetry of the messenger Lagrangian provided that
\eqn\messparii{
\CM^\dagger = U^\dagger \CM \tilde U,\qquad (\lambda F)^\dagger = U^\dagger (\lambda F)\tilde U
}
and it forces the dangerous hypercharge D-term contributions to the sfermion masses to vanish, since they are odd under it.  Of course, this parity is explicitly broken by the MSSM matter fields, so there will still be hypercharge D-term contributions from loops involving both MSSM and messenger fields. However, these only enter in at three-loops and higher, so they will be negligible compared to the two-loop mass-squareds shown above.

\subsec{Doublet/triplet splitting and small $\mu$}

In this subsection, we would like to analyze a more subtle effect
of doublet/triplet splitting on the MSSM spectrum, namely the
possibility of having small $\mu$ through a cancellation in the
running of $m_{H_u}^2$. This ``focussing" effect was first pointed
out in \refs{\agashe,\agashegraesser}.

To begin, let us recall that electroweak symmetry-breaking in the
MSSM specifies $\mu$ (up to a sign) in terms of the soft masses at
the electroweak scale. At large $\tan \beta$, the relation is
approximately
 \eqn\mueqn{ \mu^2 \approx -\half m_Z^2 -
m_{H_u}^2(m_{\tilde t})
 }
where the value of $m_{H_u}^2$ at $Q = m_{\tilde t}$ is
approximately given by its gauge mediation value \scalmassdt\ plus
the dominant contribution to the one-loop running coming from stop
loops:
 \eqn\mhueqn{ m_{H_u}^2(m_{\tilde t})
\approx m_{H_u}^2
   - { 3 \over 4 \pi^2} y_t^2 m_{\tilde{t}}^2
   \log { M_{{\rm mess},3} \over m_{\tilde{t}} }
 }
{}From \mueqn--\mhueqn, we see that $\mu$ will be small if a
cancellation can be arranged between the two terms on the RHS
above \refs{\agashe,\agashegraesser}. According to the general
formulae \scalmassdt,
\eqn\mhumtpropto{
m_{H_u}^2\propto {3\over4}{\alpha_2(M_{{\rm mess},2})^2\over
N_{{\rm eff},2}},\qquad m_{\tilde t}^2\propto
{4\over3}{\alpha_3(M_{{\rm mess},3})^2\over N_{{\rm eff},3}}
}
so a cancellation can occur if $N_{{\rm eff},3}\gg N_{{\rm
eff},2}$, i.e.\ if the colored and uncolored sparticle masses are
squashed together.\foot{In \refs{\agashe,\agashegraesser},
different numbers of OGM doublet and triplet messengers were put
in by hand, and additional heavy doublets and/or triplets were
included in an ad hoc fashion just for the sake of gauge coupling
unification. As we will show below, our models are more natural,
in that the R-symmetry guarantees the presence of heavy messengers
at the correct scales for unification, even when $N_{\rm eff,3}\ne
N_{\rm eff,2}$ and there is a large amount of doublet/triplet
splitting.}

We can elaborate upon this point more quantitatively. If we define
the degree of cancellation,
 \eqn\canceldef{\eqalign{ \eta
&=-{m_{H_u}^2(m_{\tilde{t}})\over
   m_{H_u}^2}\cr
  &\approx {4y_t^2\over 3\pi^2}
  {\alpha_3(M_{\rm mess,3})^2\over
  \alpha_2(M_{\rm mess,2})^2 }{ N_{\rm eff,2}\over N_{\rm eff,3}}\log { M_{\rm mess,3} \over m_{\tilde
  t}}-1
 }}
and require $0\le \eta\le 0.1$ (the lower bound is the requirement
of electroweak symmetry breaking), we obtain a range in $N_{\rm
eff,3}/N_{\rm eff,2}$ for a given set of triplet and doublet
messenger scales $M_{\rm mess,3}$, $M_{\rm mess,2}$. To illustrate
this, we have plotted in figure 1 an example of this range as a
function of $M_{\rm mess,3}$, for $M_{\rm mess,2}=500$ TeV (and
$y_t\approx 1$, $m_{\tilde t}\approx 1$ TeV). From this plot, we
can glean a few general facts about what it takes to achieve a
cancellation in the running of $m_{H_u}^2$.

\bigskip
\ifig\neffrbfig{A plot of the range of $N_{\rm eff,3}/N_{\rm
eff,2}$ where at least a one part in ten cancellation occurs in
the running of $m_{H_u}^2$. The range is plotted vs.\ the triplet
messenger scale $M_{\rm mess,3}$; the doublet scale $M_{\rm
mess,2}$ was fixed to be 500 {\rm TeV}.
}{\epsfxsize=.7\hsize\epsfbox{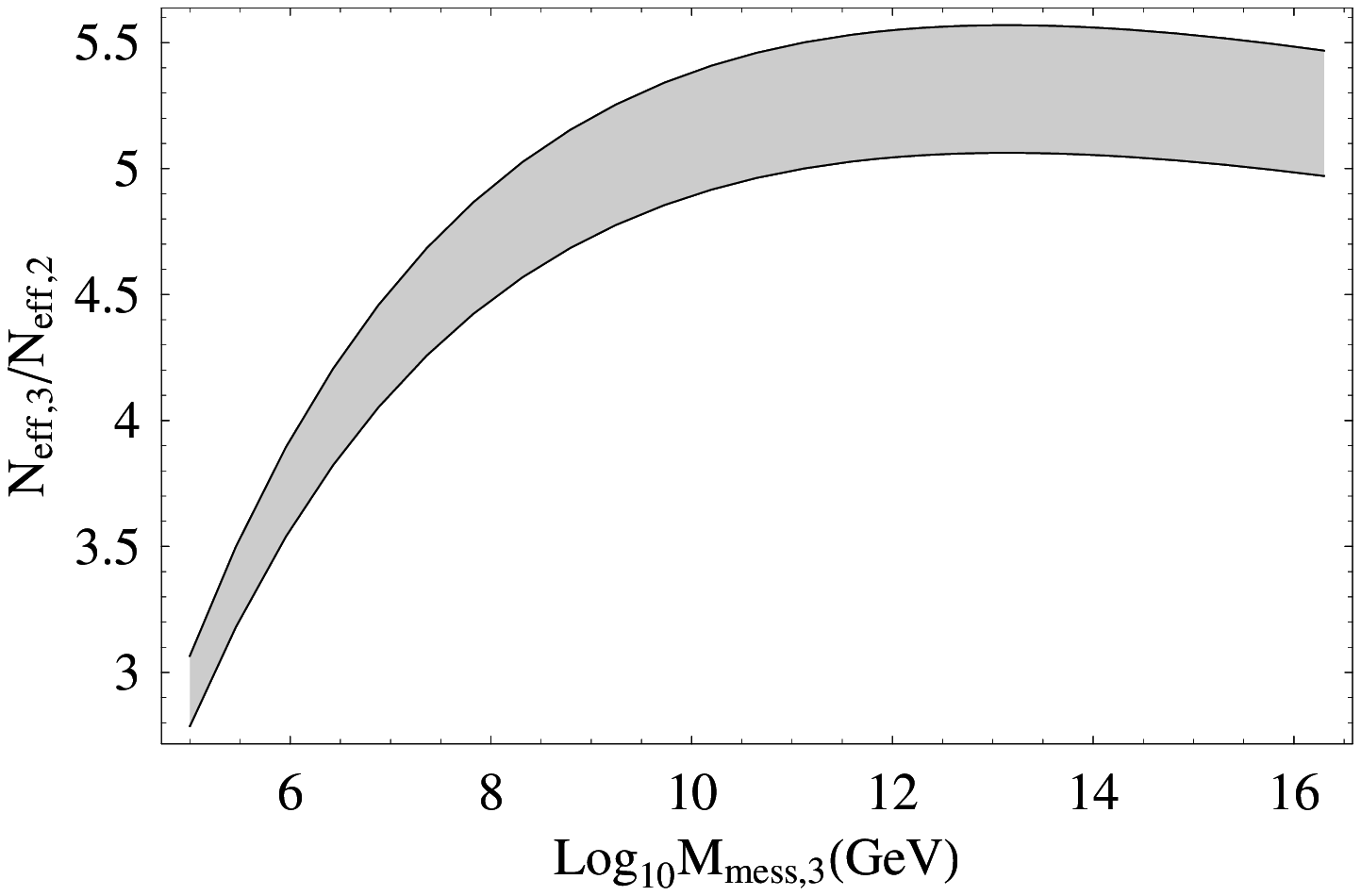}}

First, we see that $N_{\rm eff,3}/N_{\rm eff,2}$ cannot be too
large, otherwise $m_{H_u}^2(m_{\tilde t})$ is positive and
electroweak symmetry breaking does not even occur. Second, we see
that generally one needs at least three times more effective
triplet messengers than doublet messengers in order to get a
significant cancellation. A corollary of this is that in OGM one
never gets a cancellation in the running of $m_{H_u}^2$, since
there $N_{\rm eff,3}=N_{\rm eff,2}=N$. Indeed, in OGM one
typically has $|\mu|\gtrsim 1$ TeV (given the LEP bound on the
Higgs mass), and there is an absolute lower bound of $\sim 350$
GeV on $|\mu|$. By contrast, in EOGM with sufficiently many
messengers it is possible to get $|\mu|$ arbitrarily small even
while keeping fixed $m_{\tilde t}\gtrsim 1$ TeV to satisfy the LEP
bound on the Higgs mass.

Small $\mu$ is very interesting  because, among other reasons, it
implies a Higgsino-like neutralino NLSP. (Formulae for the
Higgsino fractions of the lightest neutralinos and charginos in
the small $\mu$ limit can be found in appendix C.) Although the
possibility of Higgsino NLSPs in gauge mediation has been
considered before, for instance in
\refs{\DimopoulosYQ,\BaerTX\MatchevFT\BaerPE-\CulbertsonAM}, this
scenario has not been given much attention, essentially because of
the theoretical bias from OGM where the NLSP is always either the
bino or the right-handed stau. Needless to say, the collider
phenomenology of Higgsino NLSPs can be quite different from that
of bino or stau-like NLSPs. For instance, a Higgsino NLSP will
have a suppressed branching fraction to $\gamma+\tilde G$ and
enhanced branching fractions to $h+\tilde G$ and $Z+\tilde G$.
Consequently, the classic $\gamma\gamma+\slashchar{E}_T$ channel
might no longer be the preferred discovery mode for gauge
mediation, if the Higgsino is the NLSP.

In our examples below, we will see that in models with
sufficiently many messengers, Higgsino NLSPs can occur in a wide
range of the EOGM parameter space. Therefore, we would argue that
this scenario deserves more study. Some preliminary remarks on the
phenomenology of Higgsino NLSPs are contained in appendix C.  A
detailed analysis would take us too far afield in this paper, so
we will leave this work for a future publication \ournextpaper.

\subsec{Small $\mu$ and the little hierarchy problem}

Another reason small $\mu$ is interesting is because of its
implications for naturalness and the ``little hierarchy problem."
The little hierarchy problem is usually cast in terms of the
amount of cancellation or fine-tuning required in \mueqn\ between
the supersymmetric $\mu$ parameter and the soft SUSY-breaking
$m_{H_u}^2$ parameter, in order to achieve the observed value of
$m_Z^2$. The amount of fine tuning with respect to a coupling
$\lambda$ is often quantified in terms of the Barbieri-Giudice
measure \BarbieriFN,
\eqn\BGmeasure{ \Delta_\lambda(m_Z^2) = \left|{\partial \log
m_Z^2\over\partial\log \lambda}\right| } That is,
$\Delta_{\lambda}^{-1}$ corresponds to the percent fine-tuning in
the parameter $\lambda$ required to achieve the observed value of
$m_Z^2$. For instance, the fine tuning associated with the $\mu$
parameter is
\eqn\BGmeasure{ \Delta_{\mu^2}(m_Z^2) = \left|{\partial \log
m_Z^2\over\partial\log \mu^2}\right| = {2\mu^2\over m_Z^2}
 }
As mentioned in the previous subsection, in OGM one typically has
$|\mu|\gtrsim 1$ TeV because of the LEP bound on the Higgs mass.
Thus OGM -- like much of the MSSM parameter space -- has a little
hierarchy problem in that it is fine-tuned to at least the percent
level with respect to $\mu$. (For a recent, more detailed
discussion of the fine-tuning problem in OGM, see e.g.\
\KobayashiFH.)

Now let us contrast this with the situation in EOGM. We have seen
in the previous subsection that, by having different effective
doublet and triplet messenger numbers, it is possible in EOGM to
have $\mu\sim 100$ GeV even with TeV scale stop masses. Thus, the
fine-tuning with respect to $\mu$ in EOGM can be improved to
$\CO(10\%)$ or better, and this puts us one step closer to solving
the little hierarchy problem.

Of course, the route to small $\mu$ in EOGM is through a partial
cancellation between the gauge mediation contribution to
$m_{H_u}^2$ at the messenger scale, and the radiative corrections
to $m_{H_u}^2$ coming from RG evolution down to the weak scale.
Thus one might wonder whether the reduction in fine-tuning with
respect to $\mu$ is merely being compensated for by an increased
fine-tuning with respect to other parameters responsible for the
cancellation. In fact, the situation can be better than it seems,
because the cancellation depends on $N_{\rm eff,3}/N_{\rm eff,2}$,
and if these are taking their asymptotic values at $X\to0$ or
$X\to \infty$, then they are actually {\it insensitive to the
couplings}, as noted in \Neffasymconst.

To make this a bit more precise, let us estimate the fine tuning
with respect to the other parameters of the model using
\mueqn--\canceldef\ and the Barbieri-Giudice measure. This gives
\eqn\BGmeasureother{\eqalign{
\Delta_\lambda(m_Z^2)&\approx \left|{2\lambda\over m_Z^2}{\partial
m_{H_u}^2(m_{\tilde t})\over\partial\lambda}\right|\cr
 &\approx \left|{2\lambda\over m_Z^2}\partial_\lambda \left[\left({n F\over X}\right)^2\left({3\over2}\left({\alpha_2\over
 4\pi}\right)^2 N_{\rm eff,2}^{-1}-{3\over 4\pi^2}y_t^2\times {8\over3}\left({\alpha_3\over 4\pi}\right)^2N_{\rm eff,3}^{-1}\log { M_{\rm mess,3} \over
 m_{\tilde{t}}}\right)\right]\right|
}}
If we assume that $N_{\rm eff,2}$ and $N_{\rm eff,3}$ are given by
their asymptotic values as in \Neffasymconst, then they are
essentially constants. Then the fine-tuning \BGmeasureother\ will
be negligible with respect to most of the parameters of the model;
the only ones that matter are $M_{{\rm mess},3}$, $y_t$,
$\alpha_2$ and $\alpha_3$, and $F/X$. The Barbieri-Giudice measure
for these are either the same or smaller than in a theory without
focussing. Therefore, we conclude that the overall amount of fine
tuning is reduced in these models, due to the insensitivity of the
asymptotic values of $N_{\rm eff}$ (and hence the amount of
cancellation) to the model parameters.

\subsec{Gauge coupling unification}

We have seen how doublet/triplet splitting in EOGM can have
interesting effects on the MSSM spectrum. However, all these
results would be significantly less interesting if they required
an amount of doublet/triplet splitting that ruined the successful
unification of the gauge couplings seen in the MSSM. In this
subsection, we would like to analyze this issue in detail. We will
see that because of the R-symmetry, the sensitivity of the running
of the gauge couplings to doublet/triplet splitting is
significantly reduced, meaning that it is possible to achieve all
the effects described in the previous subsections without
sacrificing unification.

To begin, let us consider the one-loop RG evolution of the gauge
couplings up to the GUT scale $m_{\rm GUT}$. After passing through
all the individual doublet and triplet messenger thresholds, one
finds that the value of the gauge couplings at $m_{\rm GUT}$
depends only on the ``average'' doublet and triplet messenger
scales,
 \eqn\Mbardef{
  \bar\CM_{2,3} \equiv (\det \CM_{2,3})^{1/N}
 }
More precisely, one finds
\eqn\grunnII{\eqalign{
 \alpha_r^{-1}(m_{\rm GUT}) &= \alpha_r^{-1}(m_Z) + {b_r  \over 2\pi} \log
 { m_{\rm GUT} \over m_Z } -
  { N \over 2 \pi}\log {m_{\rm GUT}\over \bar\CM_r} \qquad
}}
for $r=1$, 2, 3. Here $b_r = (-{33 \over 5}, -1, 3)$ denotes the
MSSM one-loop $\beta$ functions,  and $\bar\CM_1\equiv
(\bar\CM_2)^{3/5}(\bar\CM_3)^{2/5}$. Note that the first two terms
in \grunnII\ correspond to the value of the MSSM gauge couplings
at the GUT scale. As is well-known, these unify to a high degree
of precision (more on this in the next paragraph), with a common
value at the GUT scale given by
\eqn\MSSMalphagut{
 \alpha_r^{-1}(m_Z) + {b_r  \over 2\pi} \log
 { m_{\rm GUT} \over m_Z } \approx \alpha_{\rm
 GUT,MSSM}^{-1}\approx 24.3
 }
Combining \grunnII\ and \MSSMalphagut, we conclude that when
$\bar\CM_2 = \bar\CM_3$, unification occurs precisely as in the
MSSM. Furthermore, the determinant identity \detmmess\ tells us
that $\bar\CM_{2,3}=(X^nG(m_{2,3},\lambda_{2,3}))^{1/N}$, and as
we will see in the next section, the function $G$ is generally
independent of some subset of the couplings. Therefore, with this
subset of couplings, we can still achieve an arbitrary amount of
doublet/triplet splitting, while preserving the same precision of
unification seen in the MSSM.

For the sake of completeness, let us also work out how much
splitting between $\bar\CM_2$, $\bar\CM_3$ can be tolerated
without spoiling unification. A commonly used measure of
unification (see e.g. \refs{\AitchisonCF, \PeskinEZ}) is the
quantity
 \eqn\Bdef{
B \equiv { \alpha_2^{-1}(m_Z) - \alpha_3^{-1}(m_Z)  \over
   \alpha_1^{-1}(m_Z) - \alpha_2^{-1}(m_Z) }
 }
By assuming unification and running the gauge couplings down from
the GUT scale, one obtains a prediction for $B$ that can be
compared with experiment. The one-loop MSSM prediction is
$B={b_3-b_2\over b_2-b_1}={5 \over 7}$, and this agrees with
experiment to approximately 5\% accuracy, where the bulk of the
uncertainty comes from the unknown GUT and MSSM thresholds. In our
models, it follows from setting $\alpha_1(m_{\rm
GUT})=\alpha_2(m_{\rm GUT})=\alpha_3(m_{\rm GUT})$ in \grunnII\
that
 \eqn\Bus{
  B  = { (b_3 - b_2) \log \left( { m_{\rm GUT} \over m_Z} \right) +
  N\left( \log \bar\CM_3- \log \bar\CM_2 \right) \over
   (b_2 - b_1) \log \left( {m_{\rm GUT} \over m_Z} \right) -
  {2 \over 5} N\left( \log \bar\CM_3 - \log \bar\CM_2 \right) }
 }
Setting $N=0$ or $\bar\CM_3=\bar\CM_2$ in \Bus\ gives the one-loop
MSSM value. If we are to deviate no more than 5\% from this, then
we require
\eqn\unifconstr{ N\left| \log {\bar\CM_3\over\bar\CM_2} \right|
\lesssim 5 } where we have used $\log (m_{\rm GUT} / m_Z) \approx
33$. According to this inequality, the amount of splitting in the
average messenger scales that we are allowed to tolerate depends
sensitively on the messenger number $N$.  For $N=1$ we can split
the average messenger scales by as much as a factor of 100. But
for $N=5$ we can only tolerate a factor of a few. However, let us
reiterate that it is possible to have an arbitrary amount of
doublet/triplet splitting yet still keep
$\bar\CM_3\approx\bar\CM_2$, because of the determinant identity
\detmmess.

Finally, let us see what the requirement of perturbativity up to
the GUT scale looks like in these models. Taking
$\bar\CM_2\approx\bar \CM_3\equiv \bar\CM$ as required by
\unifconstr, and demanding that $\alpha_r^{-1}(m_{\rm GUT})>0$, we
find from \grunnII--\MSSMalphagut\ the following condition on $N$
and the average messenger scale:
\eqn\pertconst{
 N\left( \log { m_{\rm GUT} \over \bar\CM }\right)  \lesssim 150
}
In other words, we find the same condition as in OGM, but with the
messenger scale given by $\bar\CM$. At $\bar\CM$ = $10^3, 10^5,
10^7, 10^9$ TeV, this condition allows for $N=6,8,10,15$
messengers, respectively.

\newsec{Classification of Models}

Having deduced some general results about EOGM models, next we
would like to identify three distinct categories of models and
apply these results to each category.

\subsec{Type I: Theories with $\det\,m\ne 0$}

In these theories, it is most convenient to use a bi-unitary
transformation to go to a basis where $m$ is diagonal. In this
basis, the fields must come in pairs with R-charges
$R(\phi_i)+R(\tilde\phi_i)=2$. According to \detmmess, this means
 \eqn\ndeti{ n=0\quad {\rm
and}\,\,\,\,\, \det(\lambda X+m)=\det\,m
 }
Note that \ndeti\ necessarily implies that $\det\,\lambda=0$,
otherwise the expansion of $\det(\lambda X+m)$ in powers of $X$
would include the term $X^N\det\,\lambda$.\foot{Another, perhaps
more direct way to prove these statements is the following: in the
basis where $m$ is diagonal, let us order the $\phi_i$ fields in
increasing R-charge, $R(\phi_1)\le R(\phi_2)\le \dots\le
R(\phi_N)$. Then $\lambda$ must be {\it strictly} upper
triangular, since if $\lambda_{ij}\ne 0$, the selection rule
$0=R(\phi_i)+R(\tilde\phi_j)=R(\phi_i)-R(\phi_j)+2$ requires
$i<j$. This in turn implies all the statements above, namely that
$\det\,\lambda=0$, $\lambda X+m$ is an upper triangular matrix
with only $m$ on the diagonal, and the determinant of this matrix
is independent of $\lambda$.}

Since these models have $\det\,m\ne 0$ and $\det\,\lambda=0$, the
messengers are all stable in a neighborhood of $X=0$, but some of them
can become tachyonic at large $X$. Thus, these models have a
stable messenger sector only for
\eqn\XXmax{
|X|<X_{\rm max}
}
for some $X_{\rm max}$ which (if it is not infinite) depends on $F$ and the other
parameters of the model.
Beyond this region of stability, the model will generally
have runaway behavior, as seen in the examples of \ShihAV, and studied more
generally in \FerrettiRQ.

Because $n=0$, these models are somewhat pathological
phenomenologically: according to \mgauginogen, the gaugino masses
all vanish to leading order in $F$. In general, this leads to a
large hierarchy between the gaugino and squark masses (even when
higher order corrections in $F/M_{\rm mess}^2$ are taken into
account), which in turn exacerbates the fine-tuning problems of
gauge mediation.

The type I category comprises the bulk (if not all) of the
O'Raifeartaigh-based model-building literature. This includes some
of the early attempts
\refs{\DineGU\NappiHM\DineZB\AlvarezGaumeWY-\DimopoulosGM} at
model building with (simple variations on) the original
O'Raifeartaigh model \ORaifeartaighPR, as well as the more modern
models of \refs{\tobenomura,\IzawaGS} where many aspects of the
$n=0$ theories (including the vanishing of the gaugino masses)
were worked out in detail. More recently, there have been many
models \refs{\KitanoXG\CsakiWI\MurayamaYF\KitanoWM\KitanoWZ\DineXT\AharonyMY\AmaritiQU\AbelJX-\HabaRJ}
based on massive SQCD in the free-magnetic phase \ISSi; these also
fall in the type I category, because the O'Raifeartaigh model of
\ISSi\ is essentially a type I model. It is important to note that
in many of the models listed above,
the R-symmetry is not spontaneously broken by the
interactions of the O'Raifeartaigh model itself. As a result,
these models generally include additional interactions to break
the R-symmetry either explicitly or spontaneously. Sometimes
(e.g.\ when the R-symmetry is broken explicitly) these
interactions can give rise to leading-order gaugino masses, thus
avoiding the gaugino/squark mass hierarchy and its associated
fine-tuning problems.

This is all we would like to say about the type I models, since
these have been fairly well-explored in the literature. We would
like to emphasize that the vanishing of the gaugino masses is not
a feature of spontaneous R-symmetry breaking in general, but only
of this particular, special category of models where $n=0$. In the
vast majority of EOGM models, $n\ne 0$ and the gaugino masses are
nonzero at leading order in $F$, even with a spontaneously broken
R-symmetry. We will focus on such models in the remainder of the
paper.

\subsec{Type II: Theories with $\det\,\lambda\ne 0$}

Here it is most convenient to diagonalize $\lambda$ by a
bi-unitary transformation. Then the fields must come in pairs with
$R(\phi_i)+R(\tilde\phi_i)=0$, and so
\eqn\ndetii{ n=N \quad {\rm and}\,\,\,\,\, \det(\lambda X+m) =
X^N\det\,\lambda
 }
according to \detmmess.\foot{As in the type I models, we can see
these statements more directly by ordering the $\phi_i$ fields in
decreasing R-charge, $R(\phi_1)\ge R(\phi_2)\ge \dots\ge
R(\phi_N)$. Then $m$ must be strictly upper triangular, since
$2=R(\phi_i)+R(\tilde\phi_j)=R(\phi_i)-R(\phi_j)$ requires $i<j$.}
Note that the type II models include OGM as a special case
($m=0$), as well as all continuous deformations of OGM consistent
with the symmetries.

It is simple to sketch the messenger spectrum for the type II
models, using the fact that $\det\,\lambda\ne 0$ and $\det\,m=0$.
At large $X$, $\det\,\lambda\ne0$ implies that all the messengers
have $\CO(\lambda X)$ masses; thus
\eqn\typeIINa{ N_{\rm eff}(X\to\infty)=N }
i.e.\ the theory reduces to $N$-messenger ordinary gauge mediation
at large $X$. As $X$ approaches the origin, $\det\,m=0$ means that
some messengers have $\CO(m)$ masses while others are much
lighter, with masses that go to zero as some power of $X$.
Eventually these light messengers must become tachyonic, and from
this we learn that the type II models have a stable messenger
spectrum for
\eqn\XXmin{
|X|>X_{\rm min}
}
for some $X_{\rm min}$.

Note that these models do not suffer from the same problems as the
type I models, since $n=N\ne 0$ means that the gaugino masses are
nonzero at leading order in $F/M_{\rm mess}^2$. Thus, these models
preserve the attractive feature of OGM whereby the gaugino and
sfermion masses are generated at the same scale parametrically.

Another nice feature of this class of models has to do with
unification. According to \ndetii, $\det(\lambda X+m)$ is
completely independent of $m$. Then according to \unifconstr, this
means that $m_{2,3}$ can be split an arbitrary amount without any
effect on unification. From the low-energy perspective, this would
look like an amazing coincidence.  For instance, if we take
$\lambda_2=\lambda_3=\lambda$ and look in the regime $m_3 \ll
\lambda X \ll m_2$, the doublet and triplet messenger spectra are
completely different (following the sketch above). Nevertheless,
the R-symmetry causes the messenger masses to be arranged in such
a way that the gauge couplings still unify just as in the MSSM.

\subsec{Type III: Theories with $\det\,\lambda =\det\,m=0$}

These models have
\eqn\ndettypeiii{ 0<n<N \quad {\rm and}\,\,\,\,\, \det(\lambda X+m) =
X^n G(m,\lambda)
 }
according to \detmmess. By dimensional analysis, $G(m,\lambda)$
must depend on both $m$ and $\lambda$. Since $n\ne 0$,  the
gaugino masses are nonvanishing at leading order in $F/M_{\rm
mess}^2$, as in the type II models.

Since $\det\,m=\det\,\lambda=0$, the messenger spectrum in type
III models combines features of the type I and type II models. In
particular, there will be light messengers at both large and small
$X$ in these models. Thus these models generally have a stable
messenger sector only for $X$ in a window,
\eqn\XXmaxXmin{
X_{\rm min}<|X|<X_{\rm max}
 }
where again, $X_{\rm min}$ and $X_{\rm max}$ depend on the
parameters of the model.

Type III models yield a variety of interesting theories which (to
our knowledge) have never been discussed in the literature. One
novel feature of these models is that it is fairly common to have
$N_{\rm eff}<1$. For instance, we can see from the upper bound in
\Neffasymineqii\ that this will happen at large $X$ provided that
$n$ is sufficiently small (e.g.\ $n=1$). This is a somewhat exotic
scenario, and it allows us to achieve sfermion/gaugino mass ratios
not ordinarily seen in gauge mediation. For instance, if we keep
the sfermion masses fixed at some scale (say, to push the Higgs
mass above the LEP bound), then taking $N_{\rm eff}<1$ makes the
gauginos lighter than in OGM. Having extra-light gauginos in the
spectrum (and the gluino in particular) could be interesting, as
it could enhance sparticle production rates at the LHC relative to
OGM scenarios. In section 4.2, we will analyze in detail the
phenomenology of specific examples of type III models which have
$N_{\rm eff}<1$.

\newsec{Examples}

\subsec{Example 1: a family of type II models}

In this section, we will consider some specific examples of EOGM
models. These will serve to illustrate the general features
discussed in the previous sections.

Let us start with a simple family of type II models:
\eqn\TypeIIModel{ \CM = \lambda X+m = \pmatrix{ \lambda_1 X &
m_1 & & \cr & \ddots & \ddots & \cr & & \ddots & m_{N-1} \cr & & &
\lambda_N X }
 }
This family of models is the most general if we assign the
following R-charges to the fields: $R(\phi_i)=-2i$,
$R(\tilde\phi_i)=2i$. The form of these models is motivated by the
following considerations. In order to get the maximum effect
{}from doublet/triplet splitting, we would like for the range of
$N_{\rm eff}$ to be as large as possible. As discussed around
\typeIINa, in the type II models $N_{\rm eff}(X\to\infty)=N$, so
to maximize the spread in $N_{\rm eff}$ we would like for $N_{\rm
eff}(X\to 0)=1$. It turns out that \TypeIIModel\ is the unique
family of type II models which has $N_{\rm eff}(X\to
0)=1$.\foot{Proof: from the lower bound in \Neffasymineqi, we see
that $N_{\rm eff}(X\to0)=1$ requires $r_m=N-1$. As discussed below
\ndetii, the matrix $m$ must be strictly upper triangular in a
basis where $\lambda$ is diagonal and the R-charges are ordered
$R(\phi_i)\ge R(\phi_{i+1})$. In order for $m$ to have rank $N-1$,
it must have $m_{i,i+1}\ne 0$; then this fixes the R-charges of
the fields uniquely (up to an overall phase rotation) to be
$R(\phi_i)=-2i$, $R(\tilde\phi_i)=2i$, which in turn forces all
the other entries of $m$ to be zero.}

A simple choice for the $\lambda$'s and $m$'s that also satisfies
the requirement of messenger parity is to make them maximally uniform:
\eqn\assumecp{
 \eqalign{ \lambda_i=\lambda, \qquad m_i=m}
 }
By a rephasing of all the fields, we can always take $\lambda$, $m$, $X$ and $F$ to be real in this family of examples. (As an aside, this shows that these particular EOGM models have no dangerous CP-violating phases.) In this case, messenger parity acts
according to \messpar\ with $U=\tilde U$ given by the permutation matrix
$U_{ij} = \delta_{i,N-j+1}$, or equivalently $\phi_i\leftrightarrow \tilde \phi_{N-i+1}^*$.

The choice \assumecp\ leads to a nice simplification: by dimensional analysis, and because $X$ always appears with a
$\lambda$, $N_{\rm eff}(X,m,\lambda)$ must be a function only of
the dimensionless quantity \eqn\defx{ x = {\lambda X\over m} }
Shown in figure 2 are plots of $N_{\rm eff}(x)$ for $N= 2,3,4,5$.
(As discussed in section 3.2, the type II models have a stable
messenger sector only for $|X|>X_{\rm min}$ for some $X_{\rm
min}$. In the following we will always be implicitly taking this
bound into account.) We see that, by construction, $N_{\rm
eff}(x)$ interpolates between 1 and $N$.

\ifig\neffvsxfig{A plot of the effective messenger number $N_{\rm
eff}(x)$ vs. $x=\lambda
X/m$.}{\epsfxsize=.9\hsize\epsfbox{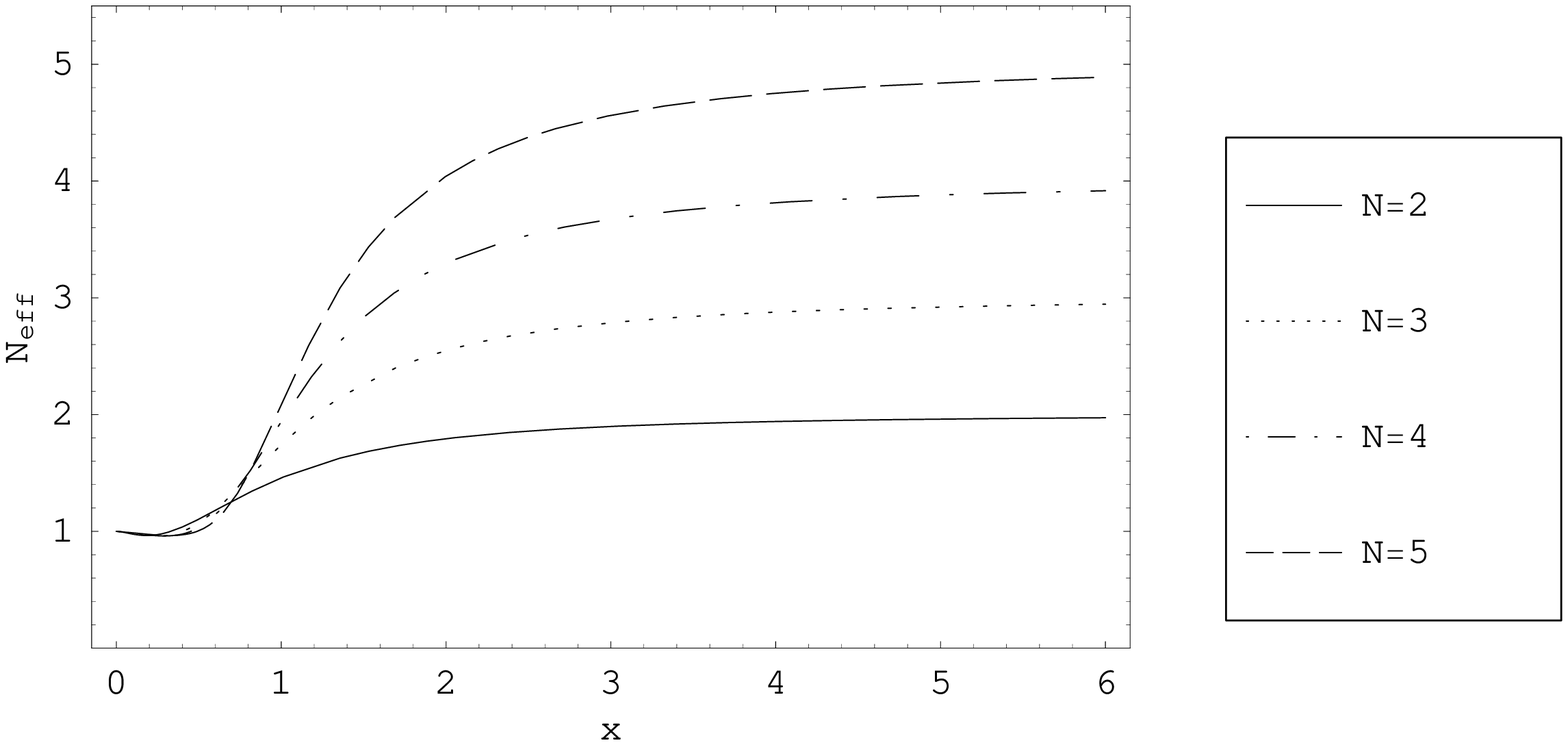}}

Now we would like to include doublet/triplet splitting and see how
it affects the phenomenology. Since unification depends only on
$\lambda_2$, $\lambda_3$ (see section 3.2), we will set
\eqn\lambdaset{
\lambda_2=\lambda_3=1
}
for simplicity. Note that when all the $\lambda$'s are the same,
the actual value of $\lambda$ is irrelevant for the current
discussion, since it always enters in the combinations $\lambda X$
and $\lambda F$.

We have generated MSSM spectra for a $N=5$ model with $m_2=2X$ and
$m_3={1\over3}X$. This implies $N_{\rm eff,2}\approx 1$ and
$N_{\rm eff,3}\approx 4.5$, as shown in \neffvsxfig. To fix the
remaining parameters ($X$ and $F$), we set $\Lambda_G=Nf/X=200$
TeV and scanned over the mass of the lightest messenger. This
choice of $\Lambda_G$ leads to stop masses around $m_{\tilde
t}\approx 1.5$ TeV and a Higgs mass around $m_{h^0}\approx 115$
GeV, which is consistent with the LEP bound, $m_{h^0} > 114.4$
GeV. Finally we have taken $\tan\beta=20$ and $\mu>0$ as a
representative choice of these parameters. The spectra are shown
plotted vs.\ the mass of the lightest messenger in figure 3. For
comparison, the spectra for an OGM model with $N=1$ and $N=5$
messengers (and all the other parameters the same) are also shown
in figure 3.

\ifig\specarrayfig{A plot of some of the MSSM soft parameters and
sparticle masses at the scale $Q=m_Z$, as a function of the
messenger scale $M$ (which we take to be the mass of the lightest
messenger). The left (middle) column is OGM with $N=1$ ($N=5$).
The right column is a model of the form \TypeIIModel\assumecp\
with $N=5$, $m_2/X = 2$, and $m_3/X = 1/3$. In all cases,
$\Lambda_G = $200 TeV.}{
\epsfxsize=1\hsize\epsfbox{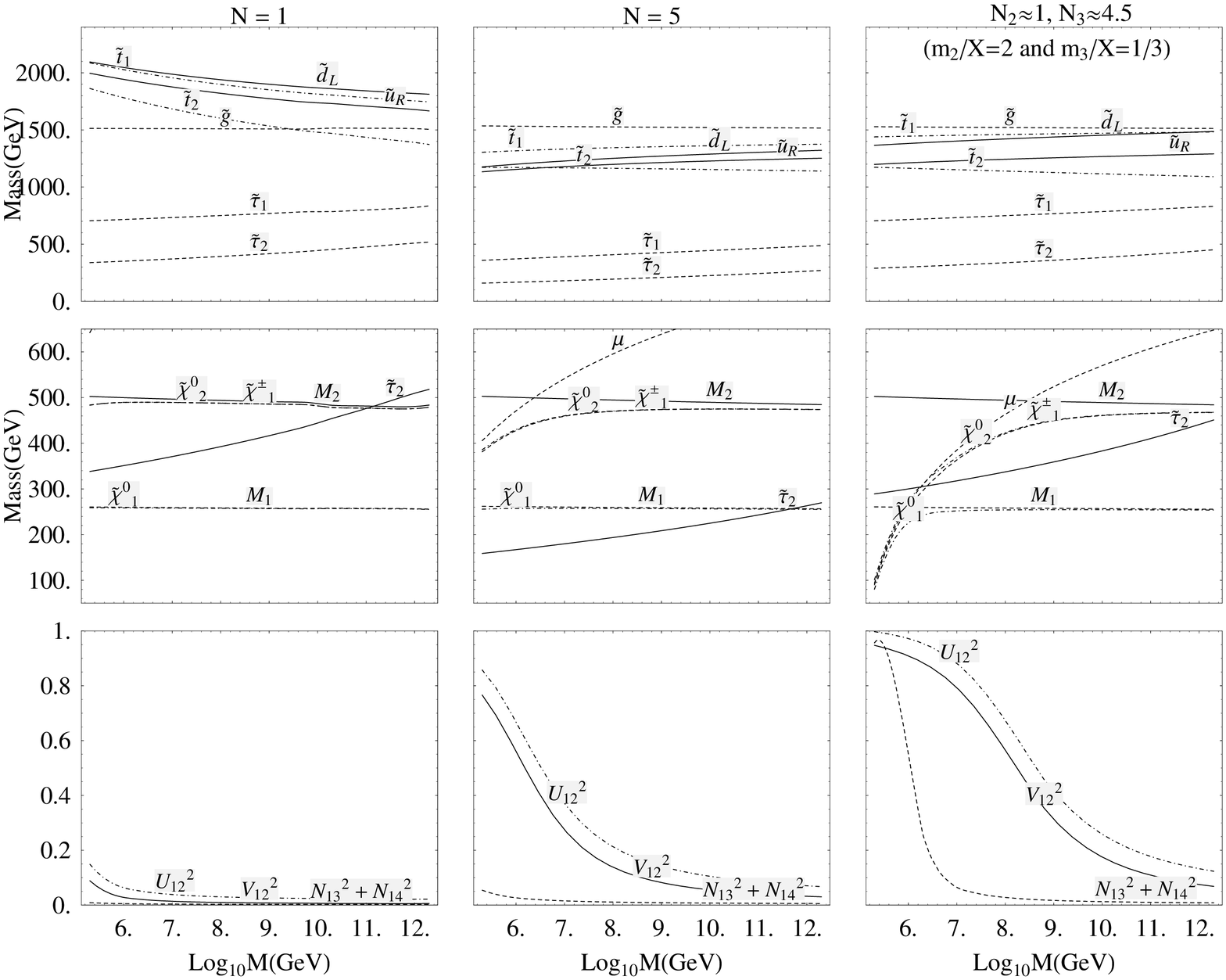}}

The spectra shown in \specarrayfig\ nicely illustrate some of the
general points made in sections 2.3 and 2.4 about the effects of
doublet/triplet splitting. For example, in the first row of
\specarrayfig\ we see that in the EOGM model the squark and
slepton masses are squashed in comparison to the $N=1$ and $N=5$
OGM models. In fact, since $\Lambda_G$ was chosen to be the same
in the three spectra, we see that the masses of colored
(uncolored) sfermions are as in the $N=5$ ($N=1$) OGM model, in
accord with the values of $N_{\rm eff,3}$ and $N_{\rm eff,2}$
respectively. By contrast, note that the values of $M_1$, $M_2$
and $m_{\tilde g}\approx M_3$ are the same between all three
models, since the R-symmetry implies that GUT relations \GUTrelmg\
always hold for the gaugino soft masses.

\ifig\muneutfig{Contour plots of $\mu$ and $N_{13}^2+N_{14}^2$ in
$\{m_2/X,m_3/X\}$ space for $N=3,4,5,6$, $M_{\rm mess} = 200
\;\rm{TeV}, 250 \;\rm{TeV}, 300 \;\rm{TeV}, 350 \;\rm{TeV}$, and
$\Lambda_G = 200 \;\rm{ TeV}$. This value of $\Lambda_G$
corresponds to $M_1=260$ {\rm GeV} and $M_2=520$ {\rm GeV}.}
{\epsfxsize=1.12\hsize\epsfbox{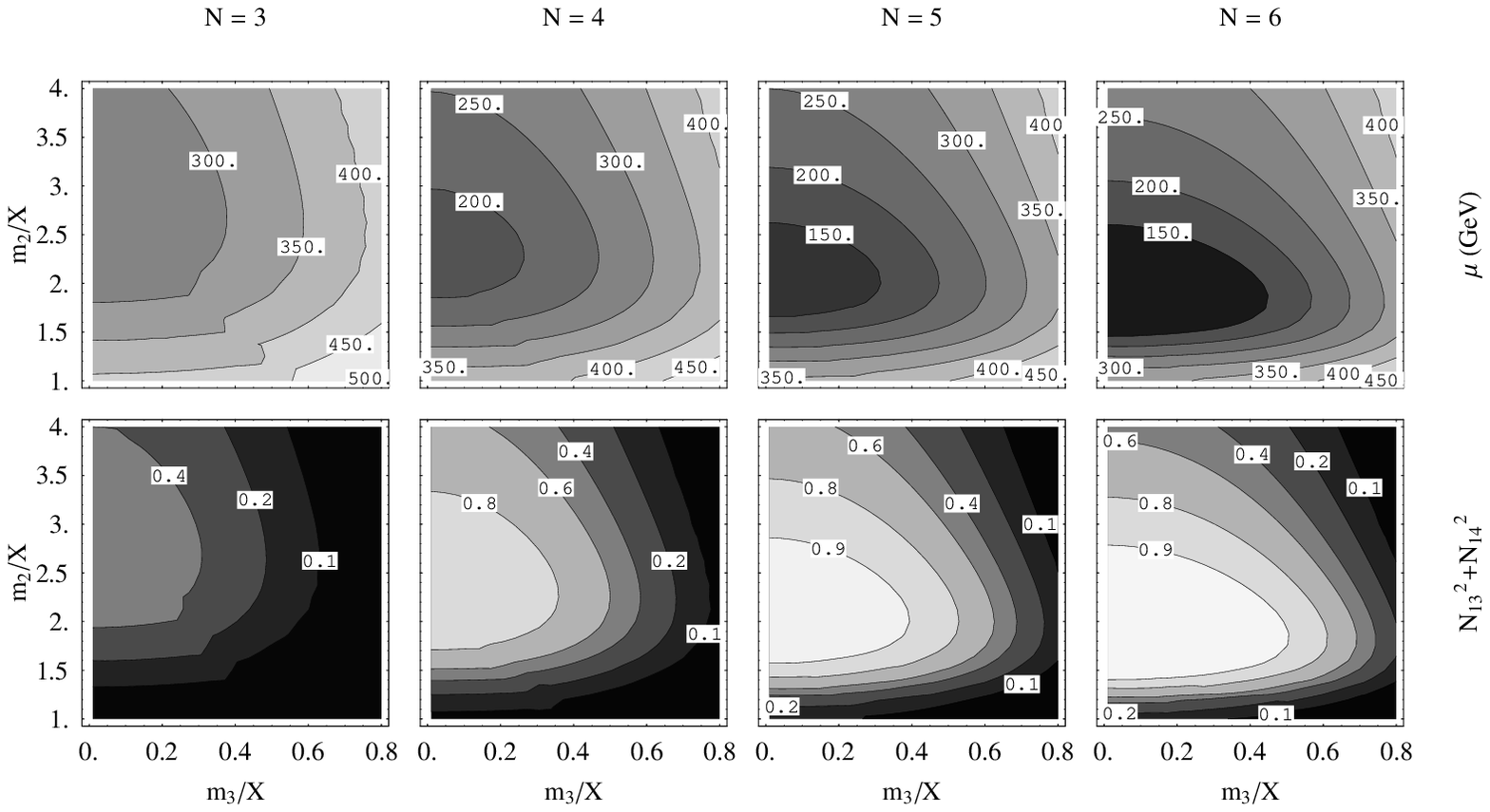}}

The second row of \specarrayfig\ contains plots of $\mu$ and the
masses of the lightest charginos, neutralinos, and stau; while the
third row contains plots of the Higgsino components of the
lightest neutralino and charginos. (For the standard definition of
the Higgsino components see appendix C.) These plots further
illustrate the consequences of doublet/triplet splitting,
specifically the dramatic effects of ``focussing'' and small $\mu$
discussed in section 2.4. To see this, consider first the $N=1$
and $N=5$ OGM spectra in \specarrayfig. These exhibit some
well-known features of OGM: $\mu$ is always large, and either the
bino ($N=1$) or the stau ($N=5$) is always the NLSP. Now contrast
this with the EOGM spectrum shown in \specarrayfig: because of the
cancellation in the running of $m_{H_u}^2$ coming from $N_{\rm
eff,3}\gg N_{\rm eff,2}$, this spectrum has $\mu\lesssim M_1$ and
a Higgsino NLSP at low messenger scales.

This point is further illustrated in \muneutfig, which contains
contour plots of $\mu$ and the Higgsino component of the lightest
neutralino vs.\ $m_2>X$ and $m_3<X$, for $N=3$, 4, 5, 6. In these
plots, we are holding fixed $\Lambda_G$ and the mass of the
lightest messenger; the parameters are chosen so that
$m_{h^0}\approx 115$ GeV. For $N \ge 4$, we see that a sizable
region of parameter space has $\mu < 200\;\rm{GeV}$ as well as an
NLSP neutralino that is more than 80\% Higgsino.

Finally, let us see how gauge coupling unification works in this
example, following the general discussion in section 2.6. Keep in
mind that throughout this subsection, we have split the doublets
and triplets in accordance with the determinant identity
\detmmess, so that unification proceeds with the same precision as
in the MSSM. In fig.\ 5, we show explicitly how the gauge
couplings run in a model with $N=3$, $m_2 = 2X$, $m_3 =
{1\over3}X$, $\Lambda_G = 200 {\rm TeV}$, and lightest messenger
mass $M_{\rm mess} =200$ TeV. For this model point, $N_{\rm
eff,2}\approx 1$ and $N_{\rm eff,3}\approx 3$, so the lightest
doublet and all three triplets contribute to the MSSM spectrum,
while the two heavy doublets essentially serve only to preserve
gauge coupling unification. The solid lines in fig.\ 5 indicate
the running of the gauge couplings up to the GUT scale; in the
magnified region around the GUT scale (inset), one can clearly see
that the gauge couplings unify to a high degree of precision. Note
that the running of the gauge couplings is very sensitive to the
location of the messenger scales, so the R-symmetry is crucial for
maintaining unification without tuning. This point is illustrated
by the dashed lines in fig.\ 5, which indicate the running of the
gauge couplings for the same model point but with the two heavy
doublet masses artificially raised by a factor of 10. In this
case, unification is already off by a significant amount, as is
clearly indicated in the inset. (Shown in the inset is also a band
obtained by varying the input value of $\alpha_3$ at $M_z$ by 5\%,
which is meant to be a rough indication of the uncertainty on
$\alpha_3$ from unknown MSSM threshold corrections and
experimental error.)

\bigskip
\ifig\typeIIunifig{The gauge couplings vs.\ RG scale $Q$ in an
$N=3$ EOGM model with $m_2=2X$ , $m_3={1\over3}X$. The vertical
lines indicate the triplet (dot-dashed) and doublet (thin dashed)
messenger masses. For comparison, the running of the gauge
couplings is also shown (thick dashed) when the two heavy doublets
are made 10 times heavier (thick solid). The inset is a
magnification of the region $Q\sim m_{GUT}$. Shown in the inset is
also a range for $\alpha_3$ corresponding to varying
$\alpha_3(M_z)$ by $\pm 5$\%.}
{\epsfxsize=0.75\hsize\epsfbox{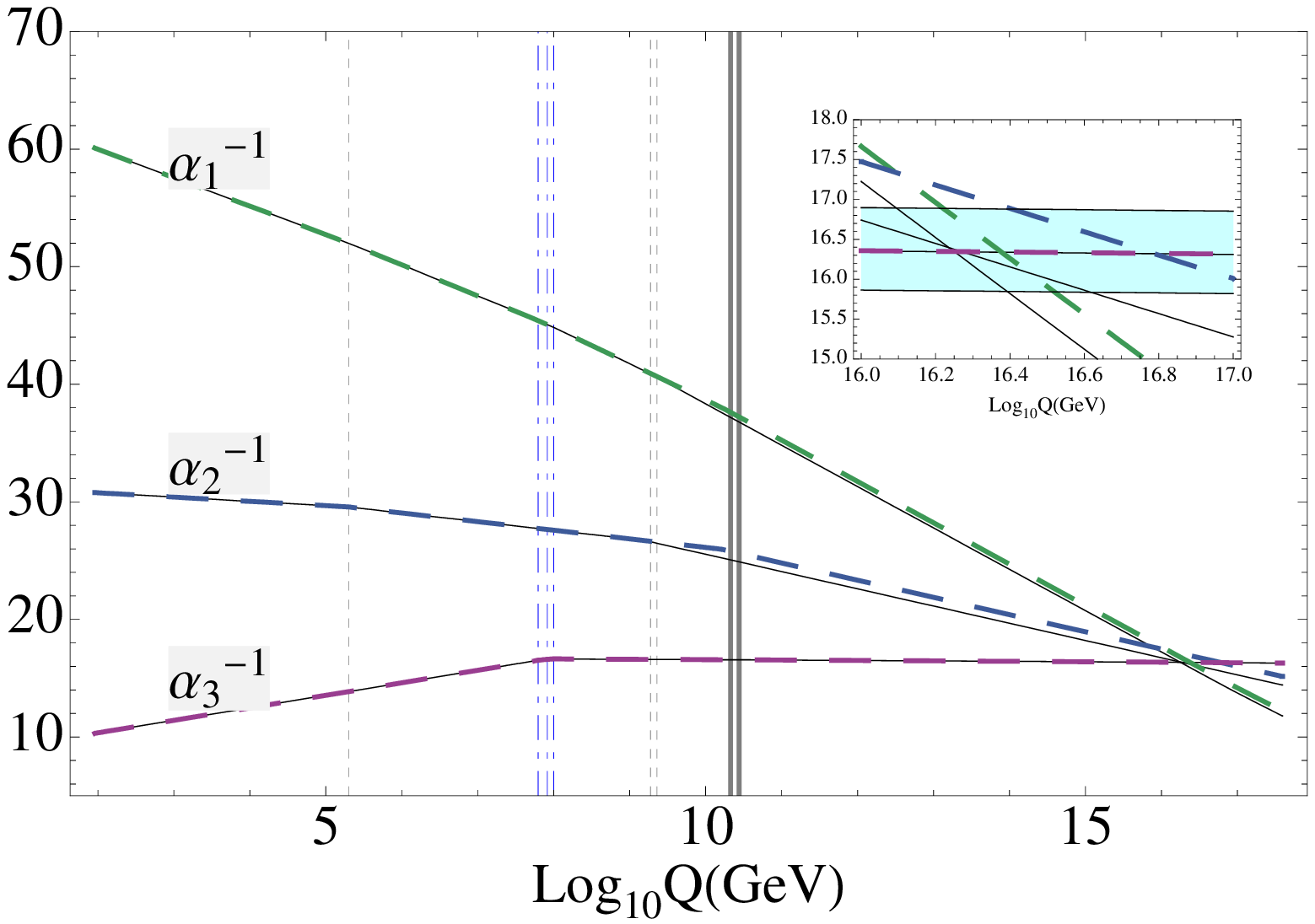}}

\subsec{Example 2: a family of type III models}

Next, let us consider a simple family of type III models which
have $n=1$ and consequently $N_{\rm eff}<1$ at large $X$. These
models are constructed by combining a single OGM messenger with an
$N-1$ messenger type I model:
\eqn\IplusII{
W = \lambda' X \phi_1 \tilde\phi_1 +
   m \sum_{i=2}^{N}  \phi_i \tilde{\phi}_i +
   \lambda X \sum_{i=2}^{N-1}  \phi_i \tilde{\phi}_{i+1}
 }
This structure can easily be enforced by proper R-charge
assignments. These models have $n=1$ because the OGM messenger
contributes $R(\phi_i)+R(\tilde\phi_i)=0$ to the formula for $n$
in \detmmess, while the $N-1$ type I messengers each contribute
$R(\phi_i)+R(\tilde\phi_i)=2$.  For the choice of couplings in
\IplusII\ (which can be taken to be real without loss of generality, as in example 1), they also have a messenger parity
defined by \messpar\ with $\tilde U=U$ and
\eqn\Umat{
U_{11}=1, \quad U_{i1}= U_{1i} = 0, \quad U_{ij} = \delta_{i-2,N-j},
\qquad (i\ge 2, j\ge 2).
}

\ifig\neffiii{$N_{\rm eff}^{-1}$ vs.\ $X$ for the model \IplusII\
for $N=3$, $4$, $5$. For simplicity, the other parameters $m$,
$\lambda$, $\lambda'$ were all set to one.
}{\epsfxsize=.75\hsize\epsfbox{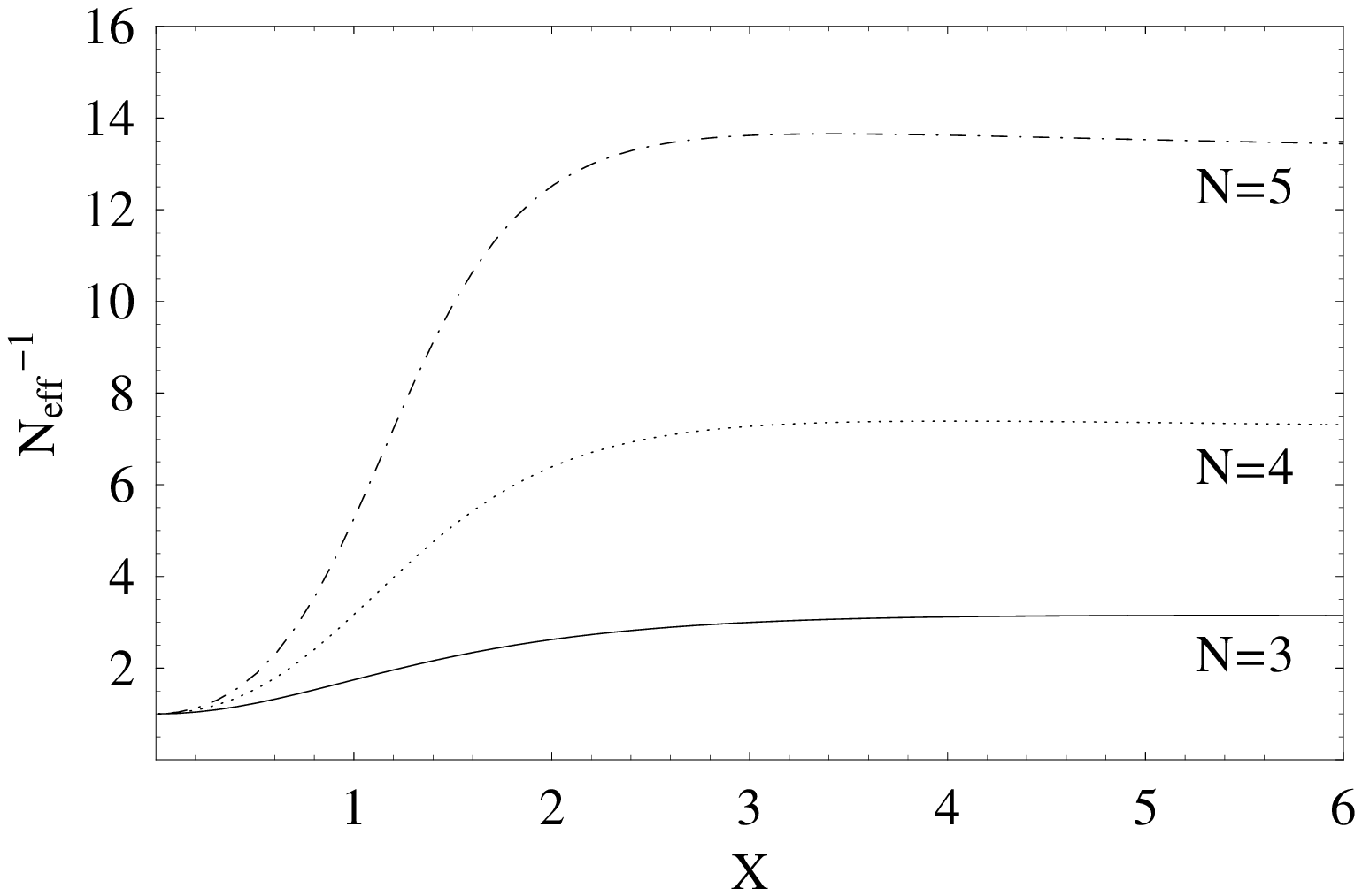}}

\noindent Since the type I piece and the OGM piece are not directly coupled through
the mass matrix, this messenger parity does not
interchange OGM messengers with type I messengers.
It is straightforward to verify
(using e.g.\ \Neffasymineqi--\Neffasymineqii) that
\eqn\NeffasymtypeIIIex{ N_{\rm eff}(X\to 0)=1,\qquad N_{\rm
eff}(X\to\infty)={1\over N-1+(N-2)^2}
} Shown in figure 6 is $N_{\rm eff}^{-1}$ vs.\ $X$ for these
models with $N=3$, 4, 5.

As discussed in section 3.3, the phenomenology of these models
with $N_{\rm eff}\ll 1$ can be quite interesting even without
doublet/triplet splitting, because when $N_{\rm eff}\ll 1$ the
gauginos are lighter than usual. Shown in the first column of
figure 7 is a sample spectrum with $N_{\rm eff}\approx 1/3$,
corresponding to an $N=3$ model with $\lambda'=\lambda=1$, $m=X /
5$, $\Lambda_G=90$ GeV, $\tan\beta=20$ and $\mu>0$. One sees from
this that the gluino mass is around 700 GeV, even though the stops
are still heavy at 1.5 TeV.

Lighter gauginos (and in particular the gluino) could mean an
enhanced rate of sparticle production at the LHC, relative to more
commonly studied OGM scenarios. Indeed, in collider studies of
gauge mediation, it is often assumed that direct gluino production
is highly suppressed relative to direct chargino and neutralino
production, because the gluino mass is generally 1 TeV or more.
However, we have seen here that in EOGM models it is possible to
have $m_{\tilde g}\sim 700$ GeV. (The gluino mass could be lowered
even further if we gave up the R-symmetry and the GUT relations.)
Even between $m_{\tilde g}\sim 700$ GeV and $m_{\tilde g}\sim 1$
TeV, the difference in the direct gluino production rate at the
LHC can be an order of magnitude or more, given the rapid fall off
of the parton luminosity functions.

By including doublet/triplet splitting, it is possible to combine
the features of type II and type III models discussed so far,
i.e.\ to have a Higgsino NLSP {\it and} a light gluino. One reason
such a scenario could be interesting is if it led to significantly
enhanced Higgs production rates at the LHC. Note that maintaining
unification is more complicated for type III models -- there is
not a clean separation in parameter space between the couplings
that enter into $\det\,\CM$ and couplings that do not. In this
example, $\det\,\CM$ depends on both $m$ and $\lambda'$, but not
$\lambda$. So if we want the same unification as in the MSSM, we
can split only $\lambda$ between the doublets and the triplets.

\ifig\tinygluino{Example spectra with and without doublet/triplet
splitting for the type III model \IplusII. The left column has
$N=3$, $\lambda'=1$, $\lambda=1$, $m=X/5$, $\Lambda_G=90$ TeV,
$\tan\beta=20$ and $\mu>0$. It shows that a $700$ GeV gluino is
possible in gauge mediation even keeping the stops heavy for the
LEP Higgs mass bound. The right plot has the same parameters,
except $N=5$, $\lambda_2=1$, $\lambda_3=1/5$. Here we see that a
light gluino and a Higgsino NLSP are simultaneously possible, at
low messenger scales.}
{\epsfxsize=.7\hsize\epsfbox{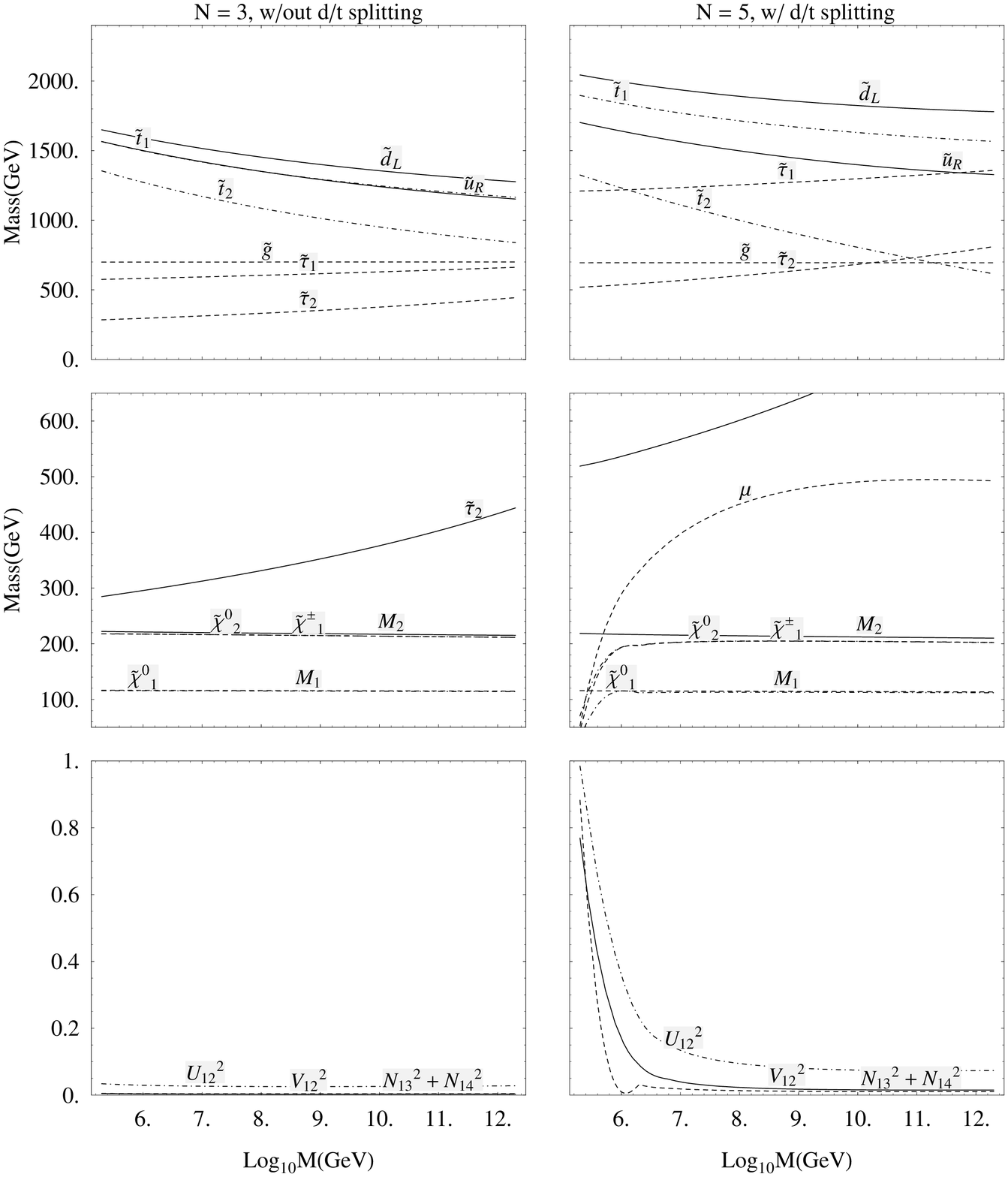}}

Shown in the second column of figure 7 is an example of a spectrum
with both a Higgsino NLSP and light gluino. The model point
corresponds to $N=5$, $\lambda_2=1$, $\lambda_3=1/5$, and all the
other parameters the same as in the previous example. Note that
$N=5$ is the minimum number of messengers required to obtain both
a light gluino and a Higgsino NLSP. The reason is that one needs
$N_{\rm eff,3}/N_{\rm eff,2}\gtrsim 3$ for small $\mu$ and
Higgsino NLSPs (see \neffrbfig), and $N_{\rm eff,3}\lesssim 1/3$
for light gluinos (as in the previous example). Together, this
implies $N_{\rm eff,2}\lesssim 1/10$. As can be seen from
\NeffasymtypeIIIex, this is possible for $N \gtrsim 5$.

We should point out that it is rather more difficult to get both a
Higgsino NLSP and a light gluino, compared to just one or the
other. One reason is simply that if $m_{\tilde g}\sim 700$ GeV,
then the GUT relations force $M_1\sim 100$ GeV, which means there
is only a very narrow window between $|\mu|=0$ and $|\mu|\sim 100$ GeV
where the NLSP has a significant Higgsino component. Another
reason is that the combination of features requires some
fine-tuning with respect to the superpotential parameters. To see
this, note that in order to have both a Higgsino NLSP and a light
gluino, we need $\lambda_2X/m$ to take an asymptotic value for
$N_{\rm eff,2}\approx 1/10$, but we need $\lambda_3X/m$ to take an
{\it intermediate} value for $N_{\rm eff,3}\approx 1/3$ (see
\neffiii). According to the discussion in section 2.5, this means
that the cancellation in the running of the Higgs mass parameter
\canceldef\ (which is controlled by $N_{{\rm eff},3}/N_{{\rm
eff,2}}$) depends sensitively on the superpotential parameters,
unlike the case when $X$ is asymptotic for both the doublets and
the triplets.

\newsec{Minimal Completions of Gauge Mediation}

\subsec{Vacuum structure}

So far, we have treated $X$ as a spurion field whose vev and
F-component are set by some undetermined hidden sector. Thus, our
approach up till this point has been analogous to most
phenomenological studies of gauge mediation, where the details of
the SUSY-breaking sector are not specified in order to be as
model-independent as possible. Now, in the last section of the paper,
we would like to go one step further and see what happens if we
require $\langle X\rangle$ to be set by the renormalizable,
perturbative dynamics of the EOGM model itself. We will see that
these dynamics {\it can} result in a viable SUSY and R-symmetry
breaking vacuum. Since the messengers play a vital role in the
SUSY breaking, this means that the models studied in this paper
can be viewed as minimal examples of direct gauge mediation.

Now let us describe our models in more detail. Given that we have
imposed $R(X)=2$ on our EOGM models, if we do not enlarge the
matter content of the theory, then the only term we can add to the
EOGM superpotential \WORii\ that is renormalizable and consistent
with the symmetries is
\eqn\deltaWcomplete{
\delta  W = F X
}
In other words, the minimal completions of our EOGM models are
just generalized O'Raifeartaigh models:
\eqn\genoragain{ W = \lambda_{ij}X\phi_i\tilde\phi_j +
 m_{ij}\phi_i\tilde\phi_j+F X}
In general, because of the R-symmetry there is a SUSY-breaking
pseudo-moduli space (i.e.\ a space of local minima of the
tree-level scalar potential) at
\eqn\pmssecv{
\phi=\tilde\phi=0,\qquad X_{\rm min}\le |X|\le X_{\rm max}
}
for some $X_{\rm min}$ and $X_{\rm max}$ (which could be zero and
infinity, respectively). In order for these models to be viable,
the one-loop Coleman-Weinberg potential must have a local minimum
on this pseudo-moduli space. Moreover, we need this minimum to
occur at $\langle X\rangle\ne0$, in order to break the R-symmetry
and give the MSSM gauginos nonzero soft masses.

We should note that, even though we are referring to these models
as generalized O'Raifeartaigh models, they generally have SUSY
vacua or runaway behavior in addition to the pseudo-moduli space
\pmssecv. (The R-symmetry, while necessary for SUSY-breaking, is
not always sufficient \NelsonNF.) Thus, the vacuum on the
pseudo-moduli space \pmssecv\ (if it exists) is only meta-stable,
and it is important to make sure that it is sufficiently
long-lived. Although we will not undertake a detailed analysis
here, on general grounds we expect that the lifetime of the
meta-stable vacuum is controlled by the small parameter $\lambda$.
This is because, using the F-terms of \genoragain\ and the
determinant identity \detmmess, one can show that the SUSY vacuum
or runaway direction in these models can only exist at
$\phi\tilde\phi\sim 1/\lambda$ and $X=0$ (or $X\to\infty$ in the
case of runaway). So the parameter $\lambda$ controls the
separation in field space between the SUSY vacuum/runaway
direction and the putative meta-stable vacuum at $\phi$,
$\tilde\phi=0$, $X\ne 0$. By making $\lambda$ small, we should be able to make
the latter parametrically long-lived.

\subsec{More on R-symmetry breaking}

It remains to determine whether, in a given model, there is a
local minimum of the Coleman-Weinberg potential with $X \ne 0$. In
\ShihAV, it was argued that this can only happen when there exists
a field with R-charge $R\ne 0$, 2. However, for technical reasons,
the argument was limited to models with $\det\,m\ne 0$. Since we
are interested in models with $\det\,m=0$ in this paper, this
argument cannot be directly applied. Nevertheless, the R-charge
condition of \ShihAV\ still seems to be true, even for models with
$\det\,m=0$. That is, regardless of whether $m$ is degenerate or
not, models where all the fields have $R=0$ or 2 never seem to
have R-symmetry breaking vacua at $X\ne 0$, while models with
exotic R-charges do. In this subsection we would like to provide
some heuristic arguments for why this should be the case.

To begin, recall that in this paper, we have been mostly
interested in the regime $\sqrt{F}\ll m$, where the approximate
formulas for the soft masses \mgauginogen--\scalmassgen\ make
sense. In this regime, the Coleman-Weinberg potential simplifies
-- it reduces to derivatives of the effective K\"ahler potential
(see e.g.\ appendix A of \ISSi\ for a detailed discussion of
this),
 \eqn\VK{ V_{\rm CW} \approx F^2\left(K_{{\rm
eff},XX^*}\right)^{-1} \sim F^2\partial^2_{XX^*}{\rm
Tr}\,\CM^\dagger\CM\log{\CM^\dagger\CM/ \mu^2}
 }
where for the sake of this heuristic discussion we are ignoring
irrelevant constants and overall normalizations. Now, it is
straightforward to apply this formula to our EOGM models and
obtain a sketch of the CW potential at large and small $X$. At
large $X$, we know on general grounds that
\eqn\VKlargeX{ V_{\rm CW}\sim F^2\log X \qquad (X\to\infty) }
i.e.\ the potential grows monotonically like a logarithm. On the
other hand, as we will now show, the behavior of \VK\ at small $X$
(by which we mean $\sqrt{F}\ll X \ll m$) depends on the R-charge
assignments of the fields.

First, let us consider a model where all the R-charges are 0 or 2.
At $X\ll m$, the fields are either heavy, with $\CO(m)$ masses, or
light, with $\CO(X)$ masses. Fields whose masses go like higher
powers of $X$ are forbidden by the R-charge assignments, as this
would require a term $\sim X^m \phi\tilde\phi$ with $m>1$ in the
effective superpotential for the light field. Now, the
contribution to the effective potential $V_{\rm CW}^{\rm (heavy)}$
from the heavy messengers must be analytic in $X$, $X^*$;
therefore, the leading dependence on $X$ in $V_{\rm CW}^{\rm
(heavy)}$ is $\CO(|X|^2)$. On the other hand, it is
straightforward to see from \VK\ that the light messengers
contribute $\sim F^2\log|X|$ to the potential. Thus the dominant
contribution at small $X$ to the potential comes from the light
messengers, and moreover, we see that it is monotonically
increasing. Given the behavior \VKlargeX\ at large $X$, the
simplest possibility is that the entire potential grows
monotonically with $X$ and has no minimum at $X\ne 0$.

Next, let us consider a model with exotic R-charged fields. Here
there can be ultra-light messengers with $\CO(X^m)$ masses with
$m\ge 2$. According to \VK, these will contribute the following to
the CW potential at small $X$,
 \eqn\LightCWPot{V_{\rm CW}^{\rm (ultra-light)}\sim
  { F^2|X|^{2m-2}} \log |X|
}
The crucial observation is that this contribution to the potential
{\it decreases} at small $X$ and eventually turns around at
intermediate $X$. Therefore, the presence of a term like
\LightCWPot\ in the potential can lead to a minimum away from the
origin.

Note that the existence of such a minimum is still not guaranteed
-- the contributions from heavier messengers of the kind discussed
above can overwhelm the effect of \LightCWPot. We will see this
happen, for instance, in some of the complete type II models to be
discussed in the next subsection.

\subsec{Type II Completions}

In this subsection and the next, we would like to study concrete
examples of complete type II and type III models. We will see that
the phenomenology of these models is more constrained than in the
previous sections, since the vev of $X$ can no longer be chosen
arbitrarily.

Consider first the type II ($\det \lambda \ne 0$) models. As
discussed in section 3.2, these models have a locally stable
pseudo-moduli space at $\phi=\tilde\phi=0$, as long as $|X|>X_{\rm
min}$ for some $X_{\rm min}$. When $|X|<X_{min}$, the potential
either runs off to infinity or to a SUSY vacuum at $X=0$, $\phi$,
$\tilde\phi\ne 0$. As discussed above, as long as $\lambda\ll 1$,
these features are well-separated from the pseudo-moduli space,
and the SUSY-breaking meta-stable vacuum (if it exists) will be
long-lived.

One nice feature of the type II completions is that as long as any
$m_{ij}\ne 0$ (respecting an R-symmetry), there must be a field
with $R\ne 0,2$ in the theory.\foot{To see this, let us again go
to a basis where $\lambda_{ij}$ is diagonal. Then
$R(\phi_i)+R(\tilde\phi_i)=0$, and any $m_{ij}\ne 0$ implies
$2=R(\phi_i)+R(\tilde\phi_j)$, so either $\phi_i$, $\tilde\phi_i$,
$\phi_j$ or $\tilde\phi_j$ must have $R\ne 0,2$.} According to the
discussion in the previous subsection, this means that the CW
potential of all these models should have a SUSY and R-symmetry
breaking minimum at $X\ne 0$, at least in some regime of
parameters. Since the type II models with $m\ne0$ comprise all the
renormalizable, R-symmetric deformations of OGM, we have
essentially shown that any such deformation of OGM -- which by
itself is an incomplete model -- will lead to a complete model of
gauge mediated SUSY breaking!

Now let us see how all this works in detail in a series of
examples, presented in order of their complexity. The simplest
example of a complete EOGM model is the $N=2$ version of the
models studied in section 4.1:
\eqn\NeqtwoOGM{
W=\lambda X( \phi_1\tilde\phi_1+\phi_2\tilde\phi_2) +
m\phi_1\tilde\phi_2+F X
 }
Notice that $\delta W=m\phi_1\tilde\phi_2$ is the only
renormalizable deformation of $N=2$ OGM consistent with {\it any}
R-symmetry (up to permutations). In this model, the boson and
fermion messenger masses can be calculated explicitly;
substituting into the approximate CW potential \VK, one finds (to
leading order in $F^2$)
\eqn\cwII{\eqalign{
 & V_{\rm CW} = {5\lambda^2 F^2\over 32 \pi^2}  V_2(x)\cr
 & V_2(x)  = - {2 \over  4 x^2+1 } + 4\log x+{ 2x^2+1 \over (4 x^2+1)^{3/2}}
    \log { 2x^2+1 + \sqrt{ 4x^2+1}\over 2x^2+ 1  - \sqrt{  4 x^2+1 }}
}} where $x\equiv \lambda X/m$. The function $V_2(x)$ is plotted in fig.\ 8;
one finds by inspection that it is minimized at $x=0.2494$.

\ifig\VCWsimpfig{The CW potential for an $N=2$ type II model (in
arbitrary units), in the small $F$ limit.
}
{\epsfxsize=.8\hsize\epsfbox{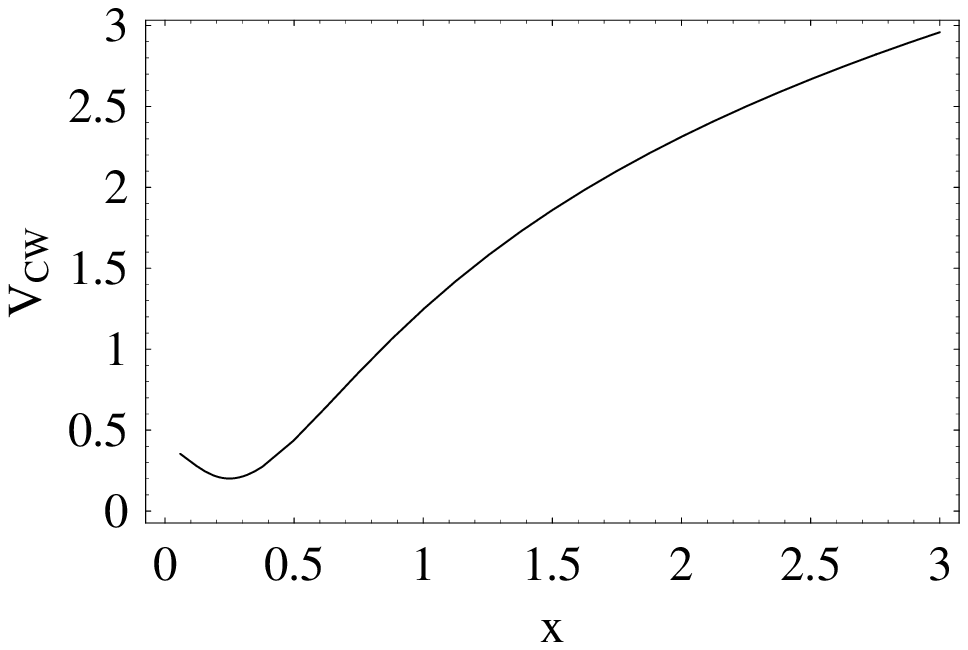}}

An analogous calculation of $V_N(x)$ for $N=3$, 4, 5 reveals that
$x$ is minimized at $(0.38,\,0.45,\,0.5)$, respectively, and there
is no minimum for $N\ge 6$.\foot{This is an artifact of choosing
$m_i = m$, $\lambda_i = \lambda$. Choosing these couplings to be
different for the different messengers can lead to a CW potential
with an R-symmetry breaking minimum.} Therefore, the $N\le 5$
models are extremely simple, complete models of direct gauge
mediation.

Consider now the effect of doublet/triplet splitting in $m_2$,
$m_3$, keeping $\lambda_2 = \lambda_3 = \lambda$ for unification.
Because of the structure of this model, the CW potential for $X$
is straightforward to compute, and it is a simple sum of
contributions from the doublet and triplet sectors:
\eqn\splitCW{\eqalign{
 V_{\rm CW}& = {\lambda^2 F^2 \over 32 \pi^2}
   \left( 2 V_N(x_2) + 3 V_N(x_2 \rho^{-1}) \right)
}}
where we have defined $x_2=\lambda X/m_2$ and $\rho = m_3/m_2$. As
described above, the first (second) term in \splitCW\ has a
minimum around $x_2\sim 1$ ($x_2\sim \rho$). Thus when $\rho \gg
1$, the second term in the potential is very flat compared to the
first, and $V_{\rm CW}$ is minimized around $x_2 \sim 1$.
Meanwhile, for $\rho \ll 1$, the opposite is true, and the minimum
of $V_{\rm CW}$ is at $x_2\sim\rho$. The upshot is that the
minimum of the potential always tracks the smaller of the two mass
parameters, i.e.\ $\langle X \rangle \sim {\rm min}(m_2,\, m_3)$.

Notice that in these examples, the vacuum always ends up at $x<1$
(or $x_2$, $x_3<1$ when there is doublet/triplet splitting). This
seems to be a general feature of these models, and there is a
simple intuitive reason for it. Namely, when $x\gtrsim 1$, the
one-loop potential is basically that of $N$ OGM messengers, i.e.\
it has no features and grows monotonically as a logarithm. Thus
the minimum of the potential, if it exists, must occur at $x<1$.

By construction (see the discussion below \TypeIIModel), the
examples considered so far have $N_{\rm eff}\approx 1$ when $x<1$.
In order to build models with $N_{\rm eff}>1$, we need to take
$r_m<N-1$, i.e.\ there must be some number of OGM messengers. If
for some reason we want to maximize $N_{\rm eff}(x\rightarrow 0)$,
then there should be as many OGM messengers as possible.

Thus we are led to a model that is the sum of a two messenger type
II model and $N-2$ OGM messengers: \eqn\hybridtypeII{ W = \lambda
X ( \phi_1 \tilde{\phi}_1 +\phi_2 \tilde{\phi}_2 )
  + m \phi_1 \tilde{\phi}_2  + \lambda'
X \sum_{i=3}^N \phi_i \tilde{\phi}_i + F X
}
This form of the superpotential could be enforced the R-symmetry,
or by a ${\Bbb Z}_2\times {\Bbb Z}_2$ symmetry that
acts separately on the OGM and the type II messengers. Note that this model
has a messenger parity symmetry which is simply the product of the separate
messenger parities of the type II and the $N-2$ OGM models. In this
model, the lower bound in \Neffasymineqi\ implies that
\eqn\Neffapprox{
N_{\rm eff}\gtrsim{N^2\over N+2}
}
when $x\lesssim 1$. So a minimum of the CW potential, if it
exists, is guaranteed to have $N_{\rm eff}>1$. In these models,
the CW potential takes the form (again at small $F$)
\eqn\dmcw{
V_{\rm CW} =  {5F^2 \over 32\pi^2} \left( \lambda^2 V_2(x) +
     2 (N-2) \lambda'^2 \log x \right)
 }
so the condition for the existence of a minimum is
\eqn\lambdareq{
 \lambda' \ll \lambda
 }
Otherwise, the contribution from the type II messengers (which has
a minimum at $x\approx 0.25$) will be overwhelmed by the
monotonically growing contribution from the OGM messengers.

Finally, in order for doublet/triplet
splitting to lead to $N_{\rm eff,3}\ne N_{\rm eff,2}$, we need to
construct a model that interpolates between \hybridtypeII\ and the higher $N$ generalizations of \NeqtwoOGM, while remaining consistent with messenger
parity.  We can achieve this with the following model:
\eqn\hybridtypeIIii{
W =  \lambda X ( \phi_{-1} \tilde{\phi}_{-1} +\phi_1 \tilde{\phi}_1 ) +
m \phi_{-1} \tilde{\phi}_1
 + \lambda' X \sum_{i=2}^{N \over 2} (\phi_i \tilde{\phi}_i
  +\phi_{-i} \tilde{\phi}_{-i})
  + \delta m \sum_{i=1}^{{N \over 2} -1} (\phi_i \tilde{\phi}_{i+1}
  +\phi_{-i} \tilde{\phi}_{-i-1}) + F X
}
where to maintain messenger parity, we have coupled the
two-messenger
type II model
in a symmetric way to two $(N-2)/2$-messenger models.  When $\delta m\to 0$, this model reduces to \hybridtypeII, and when $\delta m\to m$ this model becomes the higher $N$ generalization of \NeqtwoOGM\ (albeit with split $\lambda$, $\lambda'$). Note that
this particular interpolating model only works if the total number of messengers is
even.
%\foot{In order to get an odd number of messengers, one could couple
%a three-messenger type II model to two additional models, but this
%has an upper bound of $N_{\rm eff} \lesssim N-2$ when $x\lesssim 1$
%according to \Neffasymineqi\ and therefore does
%not lead to sufficiently large $N_{\rm eff,3}/N_{\rm eff,2}$.}
By having different $\delta m$ for the doublets and triplets, we
can make $N_{\rm eff,3}\gg N_{\rm eff,2}$ and obtain all the
exotic phenomenology (Higgsino NLSP, small $\mu$, etc.) discussed
in the previous sections, all within the context of a complete
model. To illustrate this, we have generated in figure 9 contour
plots of $\mu$ and the Higgsino component of the lightest
neutralino, for models of the form \hybridtypeIIii\ with $N=4$,
 6; $\lambda'=\lambda /10$ (to satisfy \lambdareq); and $\delta
m_2= m_2/10$, $\delta m_3=0$ so that $N_{\rm eff,3}$ is given by
\Neffapprox\ and $N_{\rm eff,2}\approx 1$. These contour plots are
scanned over $m_2/X$ and $m_3/X$, again treating $X$ as a free
parameter. The special case where $X$ is determined by the
Coleman-Weinberg potential is indicated by the solid line in
figure 9.

\ifig\mucompletefig{Contour plots for $\mu$ in $\{m_2/X,m_3/X\}$
space for $N=4,6$, $M_{\rm mess} = 200 \;\rm{TeV}, 400
\;\rm{TeV}$, and $\Lambda_G = 200 \;\rm{TeV}, 300 \;\rm{TeV}$.
Here $\lambda'=\lambda /10$, $\delta m_2=m_2 /10$, and $\delta
m_3=0$. The solid line denotes the values of $\langle X\rangle$
corresponding to the complete model.}
{\epsfxsize=0.6\hsize\epsfbox{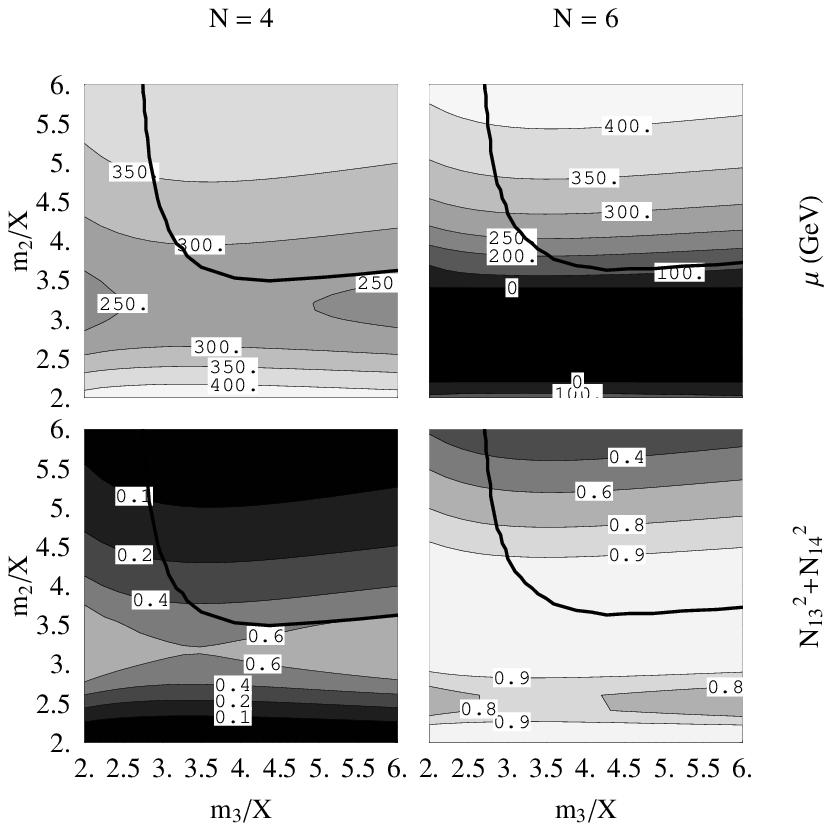}}

Let us conclude this subsection with a short summary of our
results so far. First, we have argued that the type II EOGM models
lead naturally to extremely compact, complete models of direct
gauge-mediated SUSY-breaking. We have also seen that the simplest
models have $N_{\rm eff} \approx 1$ and are largely insensitive to
doublet/triplet splitting. So in a sense, these features could be
viewed as generic predictions of these minimal models. Finally, we
constructed complete models with $N_{\rm eff}>1$ and $N_{\rm
eff,3}\gg N_{\rm eff,2}$, using the more complicated setups
\hybridtypeII\ and \hybridtypeIIii. The latter models are rather
contrived,\foot{In particular, why should $\delta m_2 \ne 0$ while
$\delta m_3 = 0$? Note that this question is similar to the
standard Higgs doublet/triplet splitting problem, with the role of
doublets and triplets reversed. There have been many ideas on how
to solve the Higgs doublet/triplet splitting problem (for a nice
overview, see \pdgRaby), and perhaps some of these ideas can be
applied here.} and they are only intended to be existence proofs,
showing that the exotic phenomenology discussed in previous
sections is possible within the space of these minimal completions
of gauge mediation.

 \ifig\VCWtypeiii{
A plot of the CW potential for the complete $N=4$ type III model
discussed in the text, with $\lambda'=0.15$, $\lambda=1$,
$m=0.1M$, and $F=10^{-4}M^2$.
}{\epsfxsize=.65\hsize\epsfbox{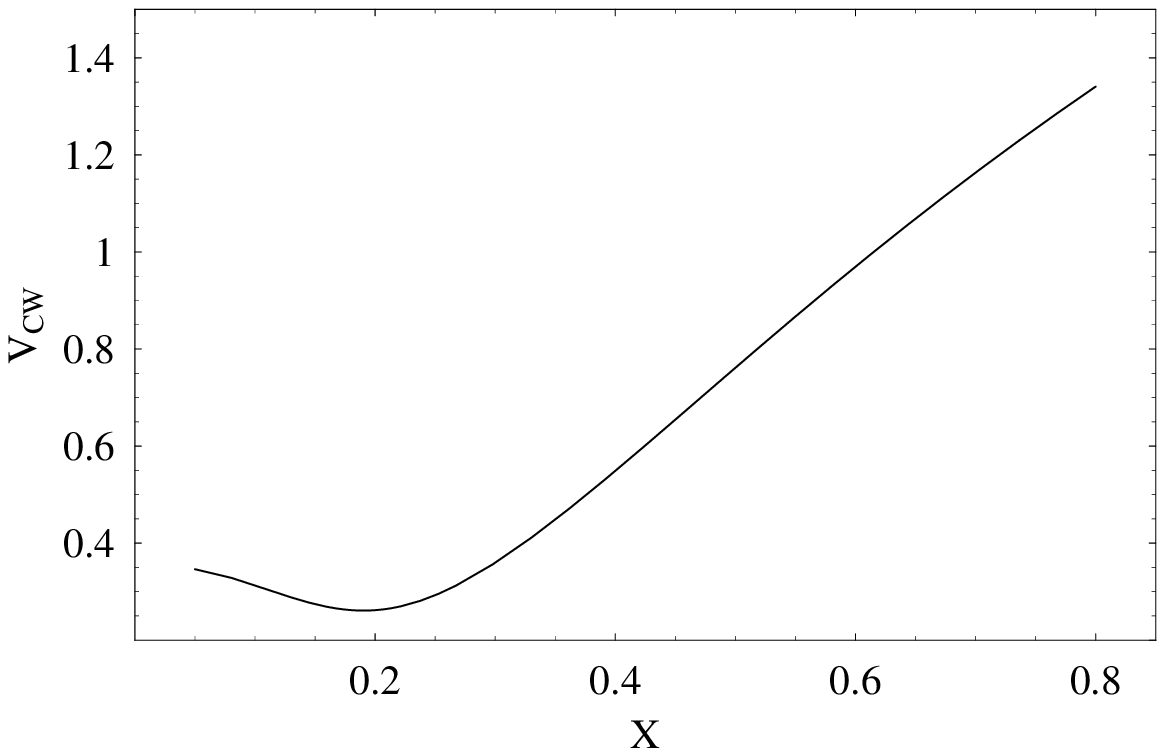}}

\subsec{Type III completions}

We would also like to explore completions of type III models.  As
we have discussed, the most interesting effects of type III models
occur when $n=1$, since this allows for the smallest possible
$N_{\rm eff}$. Thus, we will focus on completions of $n=1$ models
in this subsection. One can show that theories with $n=1$ always
contain a supersymmetric vacuum at $X= 0$, $\phi$, $\tilde\phi\ne
0$. As in the previous subsections, we will always assume that
this SUSY vacuum is sufficiently far away from the SUSY-breaking
pseudo-moduli space, so that the meta-stable vacuum (if it exists)
is long-lived.

It is straightforward to take the models \IplusII\ discussed in
section 4.2 and use them to build complete $n=1$ models with
exotic phenomenology. Recall that these models were combinations
of type I models and OGM messengers. In this section, we will
focus on a model of the form \IplusII\ with $N=4$ messengers,
\eqn\typeIIIsimp{
\lambda X+m = \pmatrix{ \lambda' X & 0  & 0 & 0 \cr
    0 & m   & \lambda X  & 0 \cr
    0 & 0 & M & \lambda X  \cr
    0 & 0 & 0 & m }
}
which respects the same messenger parity of \IplusII\ even with
$m\ne M$.
The CW potential for this model splits into a potential for the
OGM messenger and a potential for the type I model; at small $F$
this is given by
 \eqn\VCWtypeIII{ V_{\rm CW} ={5\lambda'^2  F^2  \over 16 \pi^2}
\log X + V_{\rm CW}^{\rm(type I)}
}

\ifig\speccompleteiii{ A sample spectrum for the complete type III
model with doublet/triplet splitting discussed in the text. The
parameters are as in the previous figure, but with
$\lambda_2=1.75$ and $\Lambda_G=115$ TeV.
}{\epsfxsize=.35\hsize\epsfbox{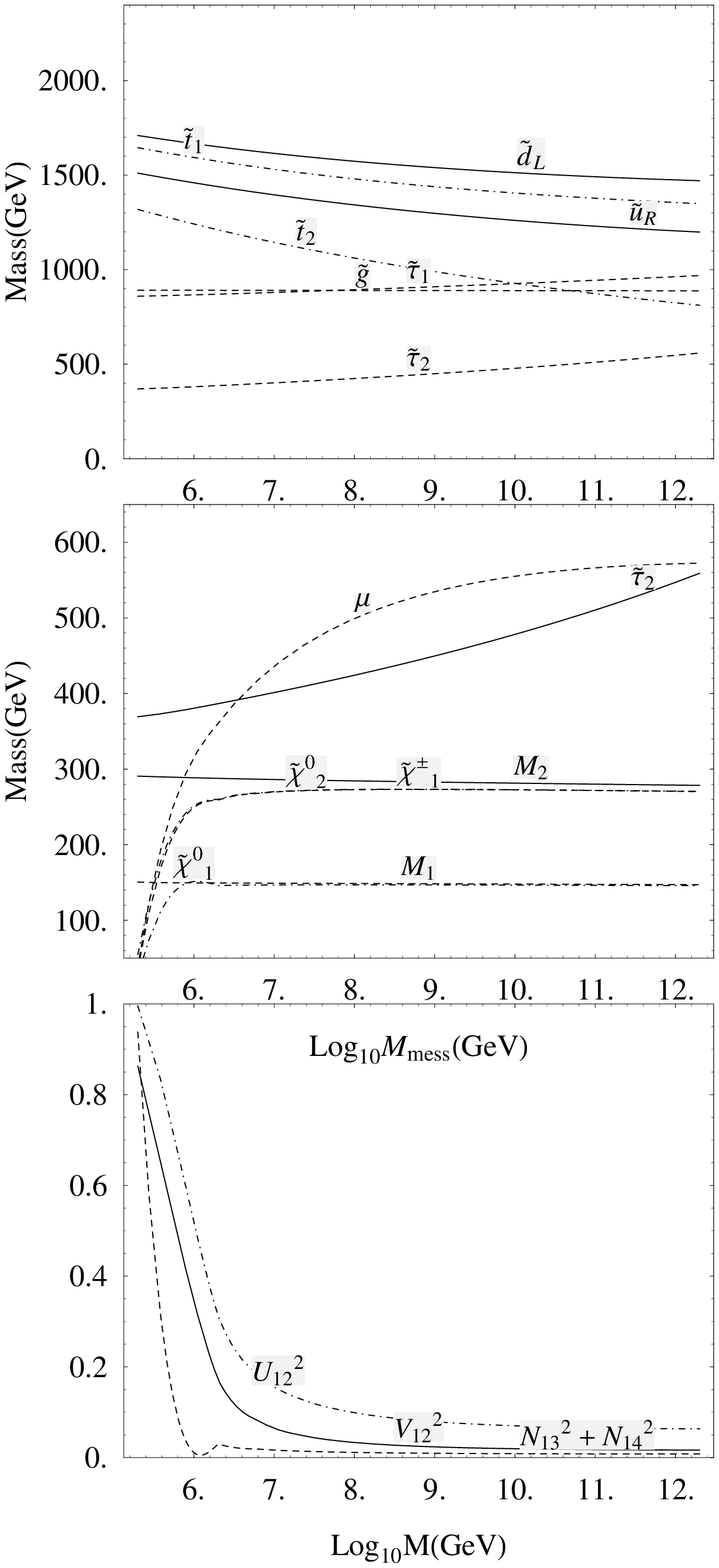}}

\noindent The type I model is precisely the one discussed in
\ShihAV; thus, we know that it has a minimum at $X\ne 0$ when
$m \ll M$. In order for the OGM messenger not to
destabilize this vacuum, we must also require $\lambda'\ll
\lambda$. An example of a potential with an R-symmetry breaking
minimum is shown in \VCWtypeiii; here we have chosen
$\lambda'=0.15$, $\lambda=1$, $m=0.1 M$, and
$F=10^{-4}M^2$.

With doublet/triplet splitting, it is possible to obtain complete
models whose spectra contain light gluinos, as well as small $\mu$
and Higgsino NLSPs. As discussed in section 4.2, we can split
$\lambda$ (but not $\lambda',m$ or $M$) between the doublets and
triplets without affecting unification. An example spectrum with
split $\lambda$'s is shown in \speccompleteiii; here the scale is
set with $\Lambda_G=115$ TeV, and the parameters are the same as
those in \VCWtypeiii, except $\lambda_2=1.75$. Note that for this
choice of parameters, $N_{\rm eff,3}=0.6$ and $N_{\rm eff,2}=0.2$.
Fig.\ 11 shows that it is possible to get gluino masses lighter
than 1 TeV, as well as Higgsino NLSPs at low messenger scales, in
a complete type III model.

\bigskip

\noindent {\bf Acknowledgments:}

First and foremost, we would like to thank M.~Dine for enlightening discussions, correspondence, and comments on the draft. We would also like
to acknowledge N.~Arkani-Hamed, P.~Meade, R.~Rattazzi for
useful discussions and N.~Seiberg for useful discussions and
comments on the draft. The research of LF is supported in part by
NSF Graduate Research Fellowship. The research of CC is supported
in part by DOE grant DE-FG02-91ER40654. The research of DS is
supported in part by DOE grants DE-FG02-91ER40654 and
DE-FG02-90ER40542. Any opinions, findings, and conclusions or
recommendations expressed in this material are those of the
author(s) and do not necessarily reflect the views of the National
Science Foundation.

\appendix{A}{The Messenger Mass Matrix}

\subsec{Determinant Identity}

Here we will prove that the R-symmetry selection rules \Rselintro\
imply the identity \detmmess:
 \eqn\detmmessrepeat{
  \det\, \CM = X^n G(m,\lambda),\qquad {\rm
where}\,\,\, n = {1 \over R(X)} \sum_{i=1}^N
\left( 2- R(\phi_i) - R(\tilde\phi_j) \right),
 }
To begin, recall the definition of the determinant \eqn\deteq{
\det\CM = \sum_{\sigma \in S_N} {\rm sgn} (\sigma)
    \CM_{1, \sigma(1)} \CM_{2, \sigma(2)} \dots \CM_{N, \sigma(N)}
 }
where $S_N$ is the degree $N$ permutation group. Now consider any
nonvanishing term in the sum \deteq, and define
\eqn\Tidef{
T_{i,\sigma} \equiv R(\phi_i) + R(\tilde\phi_{\sigma(i)})
 }
{}From the R-symmetry, $\CM_{i, \sigma(i)}$ vanishes unless
$T_{i,\sigma} = 2-R(X)$ or 2. Furthermore, if $T_{i,\sigma} = 2-R(X)$
($T_{i,\sigma} = 2$) then $\CM_{i, \sigma(i)}$ is proportional to
$X$ (a constant). Therefore, the nonvanishing term in question is
a monomial in $X$, of degree
\eqn\Xpower{
 n=\sum_{i=1}^N {\left( 2-T_{i,\sigma}\right)  \over R(X)}
= {1 \over R(X)} \sum_{i=1}^N \left(
 2- R(\phi_i)-R(\tilde\phi_{\sigma(i)})\right)  =
 {1 \over R(X)} \sum_{i=1}^N \left(
 2- R(\phi_i)-R(\tilde\phi_i)\right)
}
Note that the dependence on the permutation $\sigma$ has dropped
out in the last equation because of the sum over $N$. Therefore,
every non-vanishing contribution to the determinant is
proportional to $X^n$ with the same power $n$, and this completes
the proof of \detmmess.

\subsec{Messenger spectrum and the asymptotic behavior of $N_{\rm
eff}$}

We would like to get some idea of how $N_{\rm eff}$ depends on the
parameters of the model. But first, we need to get a rough picture
of the messenger spectrum. For this purpose, the notation
introduced in \ranksdef\ will be useful:
 \eqn\rdefs{
  r_\lambda\equiv{\rm rank}\,\lambda,\qquad
 r_m\equiv {\rm rank}\,m
 }
Note that $r_\lambda+r_m\ge N$ necessarily, otherwise $\lambda
X+m$ would be degenerate.

At large $X$, $r_\lambda$ messengers have $\CO(X)$ masses. The
remaining $N-r_\lambda$ messenger masses must scale with a smaller
power of $X$, \eqn\Mscali{ \CM_i\sim {m^{n_i+1}\over X^{n_i}}
 }
where $n_i\ge 0$ and, according to the determinant identity
\detmmess,
\eqn\nisumi{
 \sum_{i=1}^{N-r_\lambda}n_i=r_\lambda-n
  }
On the other hand, at small $X$, $r_m$ of the messengers have
$\CO(m)$ masses. According to \detmmess, the remaining $N-r_m$
messengers have \eqn\Mscalii{ \CM_i\sim {X^{n_i'+1}\over m^{n_i'}}
 }
masses, with $n_i'\ge 0$ and
\eqn\nisumii{
 \sum_{i=1}^{N-r_m}n_i'=n-(N-r_m)
 }
(By successively integrating out messengers, it is straightforward
to prove that all the $n_i$ and $n_i'$ must be integers.)
Together, these identities imply
 \eqn\nineq{
    N-r_m \le  n \le r_\lambda
 }
As a check, note that this inequality is consistent with the
inequality $r_\lambda+r_m\ge N$ deduced above.

Based on this picture of the messenger spectrum, it is trivial to
derive using \Neffdef\ the asymptotic behavior of $N_{\rm eff}$ as
$X\to 0$ and as $X\to \infty$:
\eqn\Neffasym{ N_{\rm eff}(X\to
0)={n^2\over\sum_{i=1}^{N-r_m}(n_i'+1)^2},\qquad N_{\rm
eff}(X\to\infty)={n^2\over r_\lambda + \sum_{i=1}^{N-r_\lambda}
n_i^2}
 }
Note that in both the $X\to 0$ and $X\to\infty$ limits, $N_{\rm
eff}$ is invariant under any continuous deformations of $m$ and
$\lambda$ which preserve the R-charge assignments.

Finally, combining \nisumi, \nisumii\ and \Neffasym, together with
the classic RMS-AM inequality $\langle x^2 \rangle \ge \langle x
\rangle^2$, it is straightforward to show that the asymptotic
values of $N_{\rm eff}$ satisfy the bounds
\Neffasymineqi--\Neffasymineqii\ quoted in the text.

\appendix{B}{Renormalization Methodology}
In this section, we describe how the low-energy spectra exhibited
in sections 3-5 were computed, in particular the threshold
corrections and $\beta$ functions that were used to run the soft
masses from the messenger scale down to the weak scale. The
formulae for all the corrections we used are presented in
\martinvaughn\ and \matchev. Typically, 10\% accuracy in the
low-energy soft parameters would be sufficient for the level of
phenomenological detail that concerns this paper; however, we
required much better than this since one of the most significant
effects in the low-energy spectrum was a large cancellation in the
running of $m_{H_u}^2$ between the messenger and the weak scale.
We have therefore included radiative contributions in
\martinvaughn\ or \matchev\ that correct $m_{H_u}^2$ at the
percent level.

We begin by detailing the renormalization group equation (RGE)
effects. The most straightforward of these are the two-loop
$\beta$ functions, which we have only included for gaugino masses,
$\alpha_3$, $y_t$, and $m_{H_u}^2$ itself. All other $\beta$
functions are evaluated at 1-loop.

In addition, there are new RGE effects from the messengers. Below
the scale of the scale $M_{\rm mess}$ of the lightest messenger,
the RGE's are the familiar ones of the MSSM and have been worked
out explicitly many places.  However, above $M_{\rm mess}$, the
RGE's are modified. In the class of models considered in this
article, there typically appear messengers at several different
mass scales.  Above the scale of the heaviest messenger, the
lagrangian is supersymmetric, and all soft SUSY breaking terms
vanish.  In simple gauge mediation models where all of the
messengers have the same mass, the soft terms are generated only
in the low energy theory (MSSM) where the messengers have been
integrated out, and so the messengers do not contribute to the
running. With multiple messenger thresholds, however, the soft
SUSY breaking terms begin to run as soon as the heaviest messenger
is integrated out. In between messenger thresholds, the RGE's are
those of the MSSM plus a contribution from the messengers. The
contribution to the running of a scalar $({\rm mass})^2$'s for a
general (softly broken) supersymmetric theory has been worked out
in \martinvaughn. Contributions from the MSSM enter already at
one-loop and so are naively much larger than the contribution from
the messengers. However, by dimensional analysis they are
proportional to MSSM $({\rm mass})^2$, which are themselves
suppressed by $(\alpha/ 4\pi)^2$, and therefore effectively give
only a three-loop contribution to the running. Thus, the leading
contribution is at order $\CO(\alpha^2)$ and comes from the
messenger sector (eq. (2.20) in \martinvaughn ):
\eqn\dtermrunning{
{d m^2_{\tilde f} \over d \log Q} \approx \sum_{a=1}^3 8{g_a^4
\over (4 \pi)^4}
                  C_{\tilde f}^a\, {\rm Str} (S(r) \CM^2 )
}
Here, $\CM^2$ denotes the messenger mass matrix (bosons and
fermions), and ${\rm tr} (t^A_r t^B_r) = S(r) \delta^{AB}$ defines
the Dynkin index of the representation $r$.\foot{ More precisely,
$S_a(r)$ is the Dynkin index for a single messenger $\phi_r$ for
the gauge group $G_i$; when the doublets and triplets are split,
$S_1(2) = \half ({ 3 \over 5})$ and $S_1(3) = \half ({ 2 \over
5})$ are the dynkin indices for the $U(1)$ gauge group, for a
complete doublet field and triplet field respectively.}
\dtermrunning\ has no effect above the scale of the heaviest
messenger, where the supertrace theorem clearly holds, ${ \rm
Str}\, \CM^2 = 0$; or below the scale of the lightest messenger,
where the supertrace is empty. In between the heaviest and the
lightest messenger scale, however, \dtermrunning\ has an effect,
and it is typically quite significant. We therefore include this
contribution to the MSSM $\beta$ functions in between messenger
scales. Since we include all running between messenger scales, the
threshold corrections to sfermion and gaugino masses from each
messenger is evaluated with the renormalization scale equal to the
messenger's own mass.

There are also many threshold corrections from the MSSM that are
important to include.  In particular, $m_{H_u}^2(m_{\tilde t})$ is
extremely sensitive to the top Yukawa coupling and, to a lesser
extent, $\alpha_3$. MSSM threshold corrections at the weak scale
can change $y_t$ $(\alpha_3)$ by around 10\% (20\%), which in turn
corrects $m_{H_u}^2(m_{\tilde t})$ by around 50\% when there is no
focussing, and over 100\% when there is. We include corrections to
$\alpha_3$ from stop and gluino loops:
\eqn\deltaalphaIII{
  \Delta \alpha_3 = { \alpha_3 (M_Z) \over 2 \pi}
   \left[\half - {2 \over 3} \ln \left( {m_t \over M_Z} \right)
    - 2 \ln \left( {m_{\tilde{g}} \over M_Z } \right)
   - {1 \over 6} \sum_{\tilde{q}} \sum_{i=1}^2 \ln \left(
 { m_{\tilde{q}_i} \over M_Z} \right) \right]
}
\thicksize=1pt \vskip12pt
\begintable
\tstrut |  | $y_t$ | $y_b$ | $y_\tau$ | $\alpha_3$ | $\alpha_2$
 | $\alpha_1$ | $m_{H_u}^2$ | $m_{h^0}$ \crthick
 $N=5$ EOGM |
   $\beta^{(2)}$ |-0.027\% | 0.82\% | -0.26\% | -0.39\% | 0.056\% | 0.052\% | -46.\% | \
-0.033\% \cr | $\Delta_{\rm MSSM}$ |8.4\% | 9.1\% | -0.28\% |
17.\% | 2.7\% | 1.9\% | 270 \% | 0.37\% \cr | $\beta_{\rm mess}$ |
0.020\% | 0.23\% | -0.060\% | 0.0026\% | 0.018\% | 0.016\% |
-14.\% | \ -0.041\% \crthick
 $N=5$ OGM |
   $\beta^{(2)}$ |-0.044\% | 0.39\% | -0.060\% | -0.38\% | 0.015\% | \
0.011\% | -14.41\% | -0.043\%\cr | $\Delta_{\rm MSSM}$ |8.20\% |
11.38\% | -1.71\% | 16.82\% | 2.99\% | 1.60\% | \ 83.77\% | 0.65\%
\crthick
 $N=1$ OGM |
   $\beta^{(2)}$ |-0.036\% | 0.18\% | 0.0064\% | -0.37\% | 0.018\% | \
0.0032\% | -8.18\% | -0.029\%\cr | $\Delta_{\rm MSSM}$ |8.83\% |
11.23\% | -0.78\% | 18.37\% | 3.18\% | 2.20\% | \ 88.31\% |
-0.069\%
\endtable

\bigskip

\noindent The running top Yukawa gets threshold corrections from squarks and
gluino loops, as well as from neutralinos, charginos, and Higgses.
We include all threshold corrections at 1-loop to $y_t$ (eqs. D.16
and D.18 in \matchev ).

In addition, we include less significant threshold corrections to
the standard model quarks and gauge couplings. In particular, we
include all 1-loop threshold corrections to $\alpha_{1}$ and
$\alpha_2$; these can be important, because they feed into the
definition of the running Higgs vev
$v^2=2m_Z^2/4\pi(\alpha_1+\alpha_2)$, which in turn feeds into the
definition of the running top mass.

To determine the low-energy MSSM spectrum, we employ an iterative
procedure (as in standard programs, such as SOFTSUSY 2.0
\AllanachKG) whereby an initial guess at the messenger scale is RG
evolved down to the weak scale, the MSSM threshold corrections are
computed, these are used to update the high-scale boundary
conditions, and this process is repeated until it converges to
within a 2\% change in $\mu^2$. Typically, this occurs within a
few iterations.

We have checked that in the case of OGM with $N=1$ or $N=5$
messengers, our codes matches the results of SOFTSUSY 2.0 around
the electroweak scale to 3\% or better for all parameters and to
1\% or better for $m_{H_u}^2$.

The above table summarizes the effect of these corrections on
the soft masses for an EOGM point with small $\mu$. Specifically,
we have taken the type II model of section 4.1 with
$m_2=3,m_3=1/2, \lambda=1,X=1, M_{\rm mess} = 200$ TeV and
$\Lambda_G = $ 160 TeV. For comparison, the size of the effects
are shown for $N=1,5$ OGM models with the same $M_{\rm mess}$ and
$\Lambda_G$.  The number in the table is ${X_{\rm approx} - X
\over X}$ where $X$ denotes the parameter with all corrections and
$X_{\rm approx}$ omits the indicated correction. All the running
parameters are evaluated at 1 TeV. $\beta^{(2)}$ denotes the
two-loop running, and $\Delta_{\rm MSSM}$ denotes threshold
corrections from the MSSM. For EOGM, we also show the effect
($\beta_{\rm mess}$)  of running between messenger masses.

\appendix{C}{Phenomenology of a Higgsino-like NLSP}

\subsec{Masses and mixings}

Because EOGM allows for a small $\mu$ parameter, the Higgsinos can
be lighter than the gauginos, and so the NSLP can be
Higgsino-like. To see this, recall the mass matrix for the
neutralinos and charginos:
 \eqn\chimassmatrix{\eqalign{
M_{\tilde{N}}  &= \pmatrix{ M_1 & 0 & -c_\beta s_W m_Z & s_\beta
s_W m_Z \cr
  0 & M_2 & c_\beta c_W m_Z & -s_\beta c_W m_Z \cr
 -c_\beta s_W m_Z & c_\beta c_W m_Z & 0 & -\mu \cr
s_\beta s_W m_Z & -s_\beta c_W m_Z & - \mu & 0} \cr M_{\tilde{C}}
& = \pmatrix{ 0 & {\bf X}^T \cr {\bf X } & 0 } \qquad {\bf X} =
\pmatrix{ M_2 & \sqrt{2} s_\beta m_W \cr
  \sqrt{2} c_\beta m_W & \mu }
 }}
In the limit $\mu\sim m_Z\ll M_1,\,M_2$, the masses are given by:
\eqn\chimasslim{\eqalign{
 m_{\tilde N_\pm}^2 &=\mu^2 \pm {
  \mu\, m_Z^2(M_1c_W^2+M_2s_W^2)(1\mp\sin2\beta)\over M_1M_2} +\dots\cr
  m_{\tilde N_3}^2 &=M_1^2+2m_Z^2s_W^2 + {2\mu\, m_Z^2s_W^2\sin2\beta\over M_1}+\dots\cr
  m_{\tilde N_4}^2&= M_2^2+2m_Z^2c_W^2 + {2\mu\, m_Z^2c_W^2\sin2\beta\over M_2}+\dots\cr
  } }
and \eqn\chimasslimii{\eqalign{
 m_{\tilde C_1}^2 &= \mu^2-{2\mu\, m_W^2 \sin2\beta\over M_2}+\dots\cr
 m_{\tilde C_2}^2&=   M_2^2+2m_W^2+{2\mu\, m_W^2\sin2\beta\over M_2}+\dots
}}

The mass matrices \chimassmatrix\ are diagonalized by a bi-unitary
transformation $\tilde N_{i} = N_{ij}\psi_j^0$, $\tilde
C_i^+=V_{ij}\psi_j^+$, $\tilde C_i^-=U_{ij}\psi_j^-$. In the small
$\mu$ limit, the bino, wino and Higgsino components of the
lightest neutralinos and charginos are given by the formulae:
\eqn\chicomplim{\eqalign{
&\{N_{\pm1}^2,\,N_{\pm2}^2,\,N_{\pm3}^2+N_{\pm4}^2\}
 = \cr
 &\quad \left\{ {m_Z^2s_W^2(1\mp\sin2\beta)\over 2M_1^2},\,
                      {m_Z^2c_W^2(1\mp\sin2\beta)\over 2M_2^2},\,
  1-\half m_Z^2{M_1^2c_W^2+M_2^2s_W^2\over M_1^2M_2^2}(1\mp\sin2\beta) \right
  \} +\dots\cr
 &\left\{U_{11}^2,\,U_{12}^2\right\} =  \left\{ {2m_W^2c_\beta^2\over M_2^2},\,1-{2m_W^2c_\beta^2\over M_2^2}\right\}\cr
 &\left\{V_{11}^2,\,V_{12}^2\right\} =   \left\{ {2m_W^2s_\beta^2\over M_2^2},\,1-{2m_W^2s_\beta^2\over M_2^2}\right\}\cr
 }}
Thus, in the small $\mu$ limit, the lightest neutralinos and
charginos are almost completely Higgsino.

\subsec{Decay rates}

OGM has the well-known collider signature $\gamma \gamma + \slashchar{E}$
from promptly decaying binos.
The rates for this are typically enormous (there will be thousands
of such events at the LHC after only 100 pb$^{-1}$ of data), and
the SM backgrounds are virtually non-existent
\refs{\DimopoulosVZ\AmbrosanioZR\DimopoulosVA-\AmbrosanioJN}. As
such, $\gamma\gamma+\slashchar{E}_T$ offers an excellent channel
for early discovery of gauge mediation at the LHC.

In EOGM, a Higgsino NLSP can lead to a
completely different collider signature.
Because the Higgsino is the superpartner of the Higgs, which in
turn mixes with the longitudinal mode of the Z, the branching
ratio of the NLSP to these modes is larger than in OGM. The
relative decay rates of NLSP to Goldstino + boson are given by

\eqn\DecayRates{\eqalign{{\Gamma(\chi^0_1 \rightarrow
\tilde{G}Z^0) \over \Gamma(\chi^0_1 \rightarrow \tilde{G}\gamma)}
&= {\kappa_Z \over \kappa_\gamma } \left(1-{m_Z^2 \over
m_{\chi^0_1}^2}\right)^4 \cr {\Gamma(\chi^0_1 \rightarrow
\tilde{G}h^0) \over \Gamma(\chi^0_1 \rightarrow \tilde{G}\gamma)}
&= {\kappa_h \over \kappa_\gamma } \left(1-{m_h^2 \over
m_{\chi^0_1}^2}\right)^4 }}

\eqn\Kappas{\eqalign{ \kappa_\gamma &= {m_Z^4 s_W^2 c_W^2 \over 4
M_1^4 M_2^4} (1-\sin 2 \beta)^2(M_1^2 c_W + M_2^2 s_W)^2 +\dots
\cr \kappa_Z &= {1 \over 8}(1-\sin 2 \beta)+ {m_Z^2 \cos 2 \beta^2
\over 8 M_1 M_2 \mu} (M_1 c_W + M_2 s_W) +\dots \cr \kappa_h &= {1
\over 4}(1-\sin 2 \alpha)-{m_Z^2 \cos 2 \alpha \cos 2 \beta  \over
4 M_1 M_2 \mu} (M_1 c_W + M_2 s_W) +\dots, }} where $\tan 2\alpha
= (m_A^2 + m_Z^2)/(m_A^2 - m_Z^2) \tan 2 \beta$.

Thus, if there is an appreciable separation of scales $\mu,m_Z <
M_{1,2}$, then $\kappa_{Z,h} \gg \kappa_\gamma$ and the decays to
$Z$s and Higgses will dominate over the decays to photons. Note
that because of the $\beta^4$ phase space factor, the decay rate
to $Z$'s will generally be slightly larger than the decay rate to
Higgs.

\subsec{Sparticle production at colliders}

Finally, let us discuss briefly some differences between the
production of bino vs.\ Higgsino NLSPs at hadron colliders. These
will only be very preliminary remarks; a more detailed analysis
will be contained in \ournextpaper.

Assuming gluino and squark masses above $\sim 1$ TeV, the primary
sparticle production modes at the LHC will be charginos and
neutralinos produced from s-channel weak bosons. In this scenario,
an increased Higgsino component can significantly alter the
dominant production modes and cross sections.

First, let us consider the dominant production modes for bino vs.\
Higgsino NLSPs. Because the proton PDF's fall off so quickly with
energy, the dominant production channels will generally be through
the lightest modes. When $\mu\gg M_1$, $M_2$, the two lightest
neutralinos and charginos are all gaugino-like, so we are only
concerned with couplings of s-channel weak gauge bosons to
gauginos. Hence, the relevant couplings are
\eqn\OGMProduction{\eqalign{
 q\bar q \rightarrow Z \rightarrow
\tilde{C}_1^+\tilde{C}_1^-,\qquad q\bar q'\rightarrow W^\pm
\rightarrow \tilde{C}_1^\pm \tilde{N}_2.
 }}
Direct production of $\tilde{N}_1$ is suppressed because it is
bino-like, and binos are neutral under electroweak.

Now let us contrast this with the situation for Higgsino NLSPs.
When $\mu\ll M_1$, $M_2$, the two lightest neutralinos and
charginos are all Higgsino-like and are all nearly degenerate
around $\mu$. Consequently, we care about the couplings of
s-channel weak gauge bosons to Higgsinos, and the relevant
channels are:
\eqn\EOGMProduction{\eqalign{ q \bar q \rightarrow Z\rightarrow \tilde{C}_1^+
\tilde{C}_1^- ,\qquad
  q\bar q \rightarrow Z\rightarrow \tilde{N}_1 \tilde{N}_2,\qquad
 q\bar q'\rightarrow W^\pm \rightarrow \tilde{C}_1^\pm \tilde{N}_{1,2}.
  }}
The first two channels are completely analogous to the two
production channels for bino NLSP. The third channel, however, is
an extra production mode, which is made possible because the two
lightest neutralinos are nearly degenerate Higgsinos.

Finally, let us point out another difference between Higgsino and
bino NLSPs which is apparent from \OGMProduction, \EOGMProduction.
Assuming the GUT relations amongst the gaugino masses, the
dominant channels for bino NLSPs involve wino-like particles whose
masses are $\approx 2m_{\rm NLSP}$. On the other hand, for Higgsino
NLSPs the dominant channels involve Higgsino-like particle whose
masses $\approx m_{\rm NLSP}$. Thus (at fixed NLSP mass) the sparticle
production cross sections for Higgsino NLSPs are enhanced relative
to those for bino NLSPs because the produced sparticles are
lighter.

 \listrefs \end